\documentclass[lettersize,journal]{IEEEtran}
\usepackage{amsmath,amsfonts,amssymb}
\usepackage{algorithmic}
\usepackage{algorithm}
\usepackage{array}
\usepackage{textcomp}
\usepackage{stfloats}
\usepackage{url}
\usepackage{multirow}
\usepackage{tabularx,booktabs} 
\usepackage{verbatim}
\usepackage{graphicx}
\usepackage{cite}
\usepackage{makecell}
\usepackage{url}
\usepackage{relsize} 
\usepackage{color}
\usepackage{graphicx}
\usepackage{float}
\usepackage{subfigure}
\usepackage{enumitem}
\usepackage{bbm}
\usepackage{amsmath}
\makeatletter
\renewcommand{\maketag@@@}[1]{\hbox{\m@th\normalsize\normalfont#1}}%
\makeatother
\usepackage{hyperref}  

\begin{document}

\title{On CSI-Free Multi-Antenna Schemes for Massive Wireless-Powered Underground Sensor Networks}

\author{Kaiqiang~Lin,~\IEEEmembership{Student~Member,~IEEE,}
        Onel Luis Alcaraz López,~\IEEEmembership{Member,~IEEE,}
        Hirley~Alves,~\IEEEmembership{Member,~IEEE,}
        and Tong~Hao,~\IEEEmembership{Member,~IEEE}

\thanks{K. Lin and T. Hao are with the College of Surveying and Geo-Informatics, Tongji University, Shanghai, China. E-mail: lkq1220@tongji.edu.cn; tonghao@tongji.edu.cn. (\textit{Corresponding Author: Tong Hao})} 
\thanks{ O. L. A. López and H. Alves are with the Centre for Wireless Communications, University of Oulu, Finland. Email: onel.alcarazlopez@oulu.fi; Hirley.Alves@oulu.fi.}
\thanks{This work was supported in part by National Natural Science Foundation of China (No. 42074179 and 42211530077), Academy of Finland (No. 346208 (6G Flagship) and No. 348515 (UPRISING)), the Finnish Foundation for Technology Promotion, and the China Scholarship Council.}

}

\maketitle

\begin{abstract}
Radio-frequency wireless energy transfer (WET) is a promising technology to realize wireless-powered underground sensor networks (WPUSNs) and enable sustainable underground monitoring. However, due to the severe attenuation in harsh underground soil and the tight energy budget of the underground sensors, traditional WPUSNs relying on the channel state information (CSI) are highly inefficient, especially in massive WET scenarios. To address this challenge, we comparatively assess the feasibility of several state-of-the-art CSI-free multi-antenna WET schemes for WPUSNs, under a given power budget. Moreover, to overcome the extremely low WET efficiency in underground channels, we propose a distributed CSI-free system, where multiple power beacons (PBs) simultaneously charge a large set of underground sensors without any CSI. We consider the position-aware K-Means and the position-agnostic equally-far-from-center (EFFC) approaches for the optimal deployment of the PBs. Our results evince that the performance of the proposed distributed CSI-free system can approach or even surpass that of a traditional full-CSI WET strategy, especially when adopting an appropriate CSI-free scheme, applying the advisable PBs deployment approach, and equipping the PBs with an appropriate number of antennas. Finally, we discuss the impact of underground parameters, i.e., the burial depth of devices and the volumetric water content of soil, on the system's performance, and identify potential challenges and research opportunities for practical distributed CSI-free WPUSNs deployment.    
\end{abstract}


\section{Introduction}
\IEEEPARstart{W}{ireless} underground sensor networks (WUSNs) enable the real-time monitoring of underground entities for various application scenarios, such as smart agriculture, pipeline leakage detection, disaster rescue, and border patrol, through wirelessly connected underground devices~\cite{IOUTReview, LinWUSNsMag}. However, due to high attenuation in harsh underground soil, the devices in WUSNs require much more energy than in typical terrestrial wireless sensor networks, for reliable data transmissions~\cite{LinWUSNLCN}. Furthermore, it is very cumbersome to replace the batteries of devices that are directly buried underground~\cite{VuranWUSNsreview}. To overcome these challenges, the radio-frequency (RF) wireless energy transfer (WET) technology for charging WUSNs has been recently proposed, implying wireless-powered underground sensor networks (WPUSNs). Several enabling techniques for WPUSNs have been investigated in~\cite{LiuWPUSNs,LiuMIMOWPUSNs, LiuWETMag,LinBSWPUSNs}, mostly theoretical feasibility studies. Specifically, the pioneering work by Liu~\textit{et al.}~\cite{LiuWPUSNs} conceptualized a multi-user WPUSN, where the underground devices harvest energy from an above-ground power beacon (PB) through WET, while developing a user-specific time scheduling scheme for the network throughput maximization. Meanwhile, the multiple-input multiple-output (MIMO) technology was considered in~\cite{LiuMIMOWPUSNs} to further improve the WET efficiency for the strongly-heterogeneous underground environment. Therein, the authors formulated a channel state information (CSI)-based time and energy beamforming optimization problem by considering the communication reliability and diverse data traffic demands. Furthermore, the challenges of multiple-channel access for such a system are highlighted in~\cite{LiuWETMag}. More recently, the backscatter communication technology has been considered for WPUSNs in~\cite{LinBSWPUSNs}, where the authors determined the optimal time allocation for WET and backscattering in terms of the maximum throughput performance with the assumption of CSI availability.

Notably, the above CSI-based solutions are based on the assumption that accurate CSI estimates are available. Note that the CSI acquisition is commonly realized by sending training pilots and waiting for feedback from the energy-harvesting (EH) devices. However, acquiring CSI would be very difficult and costly, if not totally infeasible, for practical WPUSN applications since: 
(i) accurate CSI estimation requires a significant amount of time and energy from underground EH devices~\cite{CSIfreeMag}; 
(ii) many training pilots are needed to obtain the CSI of the whole network, especially for a massive WPUSN deployment which shall cause high collision probability, long effective delay, and unaffordable energy consumption, thereby eliminating the gains from the CSI exploitation~\cite{OnelWETReview}; and
(iii) reliable CSI feedback in WPUSNs is particularly critical considering  the harsh channel conditions in underground soil~\cite{LinLoRaWUSNs}. Consequently, efficient WET strategies without CSI are desirable for WPUSNs, especially in massive WET scenarios where the CSI acquisition costs may be substantial. 

By intelligently exploiting the broadcast nature of wireless transmissions, several promising CSI-free multi-antenna WET schemes, hereinafter referred to just as CSI-free schemes, are proposed and analyzed in~\cite{CSIfreeTcom,CSIfreeIoT, RAB} for wirelessly powering a large set of nearby EH devices in terrestrial WET networks. For instance, the ``switching antennas'' (SA) and the ``all antennas transmitting independent signals" (AA-IS) CSI-free schemes have been developed in~\cite{CSIfreeTcom} and~\cite{CSIfreeIoT}, respectively. The numerical results in~\cite{CSIfreeIoT} evinced that AA-IS is preferable for powering the devices that are close and uniformly distributed around the PB, while those devices which are far from the PB benefit slightly more from SA. On the other hand, the ``all antennas transmitting the same signal'' (AA-SS) CSI-free scheme was proposed also in~\cite{CSIfreeTcom}, and it was later optimized to provide a wider coverage through shifting the signal's phase by $\pi$ radians in consecutive antenna elements in~\cite{CSIfreeIoT}. Herein, we refer to the default and optimized approaches as AA-SS-I and AA-SS-II, respectively. However, both AA-SS schemes are only effective in powering the terrestrial EH devices deployed in a specific cluster, depending on the orientation of the PB's antenna array. To further improve the charging coverage, the authors in~\cite{RAB} proposed the ``rotary antenna beamforming'' (RAB) CSI-free scheme, which requires the PB to be equipped with a servo motor and be continuously rotating while using the AA-SS-II scheme. Remarkably, RAB enhances the RF energy available at terrestrial EH devices compared with the other state-of-the-art CSI-free schemes, i.e., SA, AA-IS, AA-SS-I, and AA-SS-II. 

However, the power consumed by the transmitter power amplifier, circuit, and base-band processing in the RF chain is neglected in the analysis of these CSI-free schemes~\cite{TxPower, Prf}. Compared with SA, AA-IS, AA-SS-I, and AA-SS-II, RAB requires an extra hardware component, i.e., servo motor, which may increase hardware complexity and energy consumption to an extent that has not yet been investigated. Therefore, the power consumption should be taken into account to fairly assess and compare the performance of these state-of-the-art CSI-free schemes. Meanwhile, since the budgeted power cannot be entirely converted into the transmit power due to the aforementioned power consumption, the incident RF energy of CSI-free schemes deteriorates and may not satisfy the RF energy requirement in the harsh underground channel. 

To improve the RF energy available at terrestrial EH devices, a distributed-PBs-based solution is recently proposed in~\cite{DistributedWET1, DistributedWET2} to improve the WET efficiency, while homogenizing the energy provided in a charging area. Herein, PBs deployment optimization is the key to ensure such a solution benefits in relatively static terrestrial WET networks, and it is evidenced in many studies~\cite{DistributedWET3, DistributedWET4, DistributedWET5, DistributedWET6, EFFCscheme}. However, most of the studies optimizing the PBs placements are based on the prior information of the devices' positions~\cite{DistributedWET3, DistributedWET4, DistributedWET5, DistributedWET6}. Although the precise devices’ positions in the terrestrial WET networks can be known from the prior deployment planning or be collected through positioning procedures, it might be challenging in underground environments as the position accuracy is poor due to the common difficulties faced by non-destructive testing technologies\cite{HaoAirGround}. Furthermore, additional energy consumed by position information acquisition would impair massive WPUSNs scenarios. The Equally-Far-From-Center (EFFC) approach without prior knowledge about devices' locations is proposed in~\cite{EFFCscheme} to find the optimal PBs deployment satisfying an energy outage constraint in terrestrial WET networks. Herein, a thorough evaluation is lacking for the energy outage probability of those state-of-the-art CSI-free schemes, let alone our considered massive CSI-free WPUSNs scenarios.   

\subsection{Motivations and Contributions}
The WPUSN scenario under consideration in this paper differs substantially from the most commonly investigated terrestrial WET networks, primarily due to the introduced underground communication channel component, which is characterized by a much higher attenuation than the in-air channel, i.e., the impact of some new factors such as the soil’s volumetric water content (VWC) and the burial depth. Therefore, herein we explore whether the state-of-the-art CSI-free schemes are potentially feasible for powering massive underground EH devices, and focus on what affects, and how to improve, the design and operation of such schemes. To the authors’ best knowledge, this work may be the first of its kind to assess the feasibility and performance of the aforementioned state-of-the-art CSI-free schemes in massive WPUSNs scenarios. Our specific contributions are summarized as follows:
\begin{enumerate}
    \item We conduct a thorough feasibility study of the state-of-the-art CSI-free schemes, i.e., AA-IS, SA, AA-SS-I, AA-SS-II, and RAB, in massive WPUSNs; and in order to overcome the low WET efficiency in WPUSNs, we advocate integrating the CSI-free schemes with a distributed WET system~\cite{DistributedWET1, DistributedWET2}, in which multiple multi-antenna PBs are distributed over the space to charge a massive number of underground EH devices. Moreover, we consider the position-aware (i.e., K-Means~\cite{Kmeans}) and the position-agnostic (i.e., EFFC~\cite{EFFCscheme}) approaches for optimizing the PBs' deployment. 
    
    \item We model practical power budgets for the proposed distributed CSI-free system. Specifically, we carefully take the power consumed by the circuit and operations into account, including the servo motor's power consumption of RAB, which is overlooked in~\cite{RAB}. It is noteworthy that contrary to [13], which introduced an ideal RAB, we consider a practical implementation of the rotation mechanism, including the duty cycle restrictions imposed by the servo motors. This allows for establishing a fair comparison among the considered CSI-free schemes.
    
    \item Through the simulation of real-world agricultural scenarios, we corroborate that the average worst-case RF energy available at underground EH devices under RAB may suffice. However, it depends on the specific number of PBs, the number of antennas, the PBs deployment approach, burial depth, and the soil's VWC. Notably, due to the motor's power consumption, the performance of RAB not always improves with the number of antennas. Still, there is an optimum number of antennas to equip in the PBs. Meanwhile, SA, AA-IS, AA-SS-I, and AA-SS-II fail to charge underground EH devices, where the performance of AA-IS, AA-SS-I, and AA-SS-II deteriorates significantly with a greater number of antennas. Moreover, we show that RAB can outperform the full-CSI strategy under strong line-of-sight (LOS) conditions. 
    
    \item We demonstrate that the performance of EFFC is close to that of K-Means, and even better when adopting SA, AA-IS, and RAB, e.g., when deploying 8 PBs for powering 64 underground EH devices randomly deployed in a 5-m radius area. This finding highlights that efficient CSI-free WET can be attained even without devices' positioning information and can facilitate the establishment of a practical distributed CSI-free system in WPUSNs.  

    \item We analyze the effects of the key underground parameters (i.e., VWC and burial depths) on the practical performance of feasible CSI-free schemes and reveal that the distributed CSI-free system is indeed feasible for friendly underground environments. However, high-VWC and high-depth conditions still require a higher power budget, more PBs to be deployed, and specialized transmit power mechanisms.  
\end{enumerate}

\begin{table}[t]
\caption{List of Symbols}
\label{Acronyms}
\centering
\begin{tabular}{p{0.07\textwidth} p{0.37\textwidth}}%
\toprule
\textbf{Symbol}                & \textbf{Definition}                                           \\  \hline
$\mathcal{M}$, $\mathrm{PB}_{m}$&Set of PBs, and the $m$th PB in the set                      \\
$\mathcal{U}$, $U_i$           &Set of underground EH devices, and the $i$th underground EH devices in the set  \\
$M$                            &Number of PBs                                                  \\
$N$                            &Number of underground EH devices                                                  \\
$Q$                            &Number of antenna elements per PB                                    \\ 
$R$                            &Radius of the service area                                    \\
$d_{u}$                        &Burial depth of the underground EH device                          \\
$\kappa$                       &Rician fading factor                                           \\
$\mathbf{h}_{m,i}^{\mathrm{los}}$             &Normalized channel vector between $\mathrm{PB}_{m}$ and $U_{i}$ \\
$\mathbf{R}$                   &NLOS channel covariance matrix                                 \\    
$\phi_{t, m, i}$               &Mean phase shift of the $t$th antenna element relative to the first antenna at the $\mathrm{PB}_{m}$ as observed by $U_{i}$\\
$\theta_{m,i}$                 &The $i$th underground EH device's azimuth angle with respect to the transmitting ULA of $\mathrm{PB}_{m}$ \\
$ \delta_{m,i}$                &Total path losses between $\mathrm{PB}_{m}$ and $U_{i}$        \\
$L_{m,i}^{a}$                  &Air attenuation between $\mathrm{PB}_{m}$ and $U_{i}$          \\    
$L_{m,i}^{r}$                  &Refraction loss between $\mathrm{PB}_{m}$ and $U_{i}$          \\ 
$L_{m,i}^{u}$                  &Soil attenuation between $\mathrm{PB}_{m}$ and $U_{i}$         \\
$f$                            &Carrier frequency                                              \\
$c$                            &Speed of light in free space                                                \\
$\tau$                         &Path-loss exponent in air                                      \\
$l_{m,i}$                      &Air propagation distance from $\mathrm{PB}_{m}$ to $U_{i}$     \\
$d_{m,i}$                      &Soil propagation distance from $\mathrm{PB}_{m}$ to $U_{i}$    \\
$\alpha$, $\beta$              &Attenuation and phase shifting constant in soil      \\
$\mu_{r}$, $\mu_{0}$, $\varepsilon_{0}$ &Soil’s relative permeability, free-space permeability, and free-space permittivity\\
$\varepsilon'$, $\varepsilon''$&Real and imaginary parts of the relative permittivity of soil    \\
$K$                            &Number of energy signals                                      \\
$s_k$, $y_{m,i}$               &Transmit signal of the $k$th $\mathrm{PB}_{m}$, and signal produced by $\mathrm{PB}_{m}$ received at $U_{i}$ \\
$\mathbf{v}_k$                 &Normalized precoding vector associated with the $k$th energy signal\\
$p_k$                          &Transmit power of the $k$th energy signal                      \\
$p_m$                          &Total transmit power of the $\mathrm{PB}_{m}$                     \\
$\xi_{m,i}$                    &The $m$th $\mathrm{PB}$'s incident RF energy available at $U_{i}$    \\
$g(\cdot)$, $\zeta$, $E_{i}$   &EH transfer function, EH efficiency, and harvested energy at $U_{i}$ \\
$T$                            &A normalized time block                                       \\
$G_i$                          &Total incident RF energy at $U_{i}$                           \\
$P_{budget}$                   &Budgeted power of each PB                                     \\
$P_{rf}$                       &Base-band processing power consumption in a RF chain          \\
$P_c$                          &Circuit power consumption                                     \\
$\eta$                         &Amplifier efficiency                                          \\
$P_{motor}$                    &Power consumption of the servo motor's operations             \\
$T_0$                          &Pulse width for the shaft at $-\pi/2$                         \\
$T_Q$                          &Pulse width for the shaft at $\pi/2$                          \\
$T_f$                          &Duty cycle of the servo motor                                 \\
$V_{supply}$                   &Supply voltage of the servo motor                             \\
$I_{work}$                     &Working current of the servo motor                            \\ 
$\gamma$                       &EH threshold                                                  \\
$L$                            &Total number of antennas in the system                    \\ 
\bottomrule
\label{Symbolstab}
\end{tabular}
\end{table}

\subsection{Organization and Notations}
The remaining part of the paper is organized as follows. Section~\ref{system} describes the system model, and Section~\ref{CSIfreeSchemes} overviews the state-of-the-art CSI-free schemes. Then, Section~\ref{DPCS} introduces the distributed CSI-free system with two deployment approaches that enables WET in massive WPUSNs. In Section~\ref{PowerBudget}, we model the power budget by considering ideal and practical systems. Section~\ref{NumRes} presents and discusses numerical results. Finally, Section \ref{Conclusion} concludes the paper and pinpoints the outlook.

\textit{Notation:} Boldface lowercase and uppercase letters represent column vectors and matrices, respectively. For instance, $\mathbf{x}=\{x_i\}$, where $x_i$ is the $i$th element of vector $\mathbf{x}$, while $\mathbf{X}=\{x_{i,j}\}$, where $x_{i,j}$ is the $i$th row $j$th column element of matrix $\mathbf{X}$. We denote a vector of ones by $\mathbf{1}$. Superscripts $(\cdot)^H$ and $(\cdot)^T$ denote the transpose and Hermitian operations, respectively, while $\mathbbm{E} [\cdot]$ denotes the statistical expectation. Additionally, $\mathbb{C}$ and $\mathbb{R}$ are the set of complex and real numbers, respectively, and $\mathbbm{i}=\sqrt{-1}$ is the imaginary unit. Meanwhile, $|\cdot|$, $\lfloor \cdot \rfloor$, and mod$(a,b)$ are the absolute, round down (floor), and modulo operations, respectively, while $\|\mathbf{x}\|$ denotes the euclidean norm of $\mathbf{x}$. Finally, $\mathbf{w} \sim \mathcal{CN}(\mathbf{0},\mathbf{R})$ is a circularly-symmetric complex Gaussian random vector with a zero-mean vector and covariance matrix $\mathbf{R}$. Table~\ref{Symbolstab} lists the main symbols used throughout the article. 

\section{System Model} \label{system}

\begin{figure}[t]
    \centering
    \includegraphics[width=3.45in]{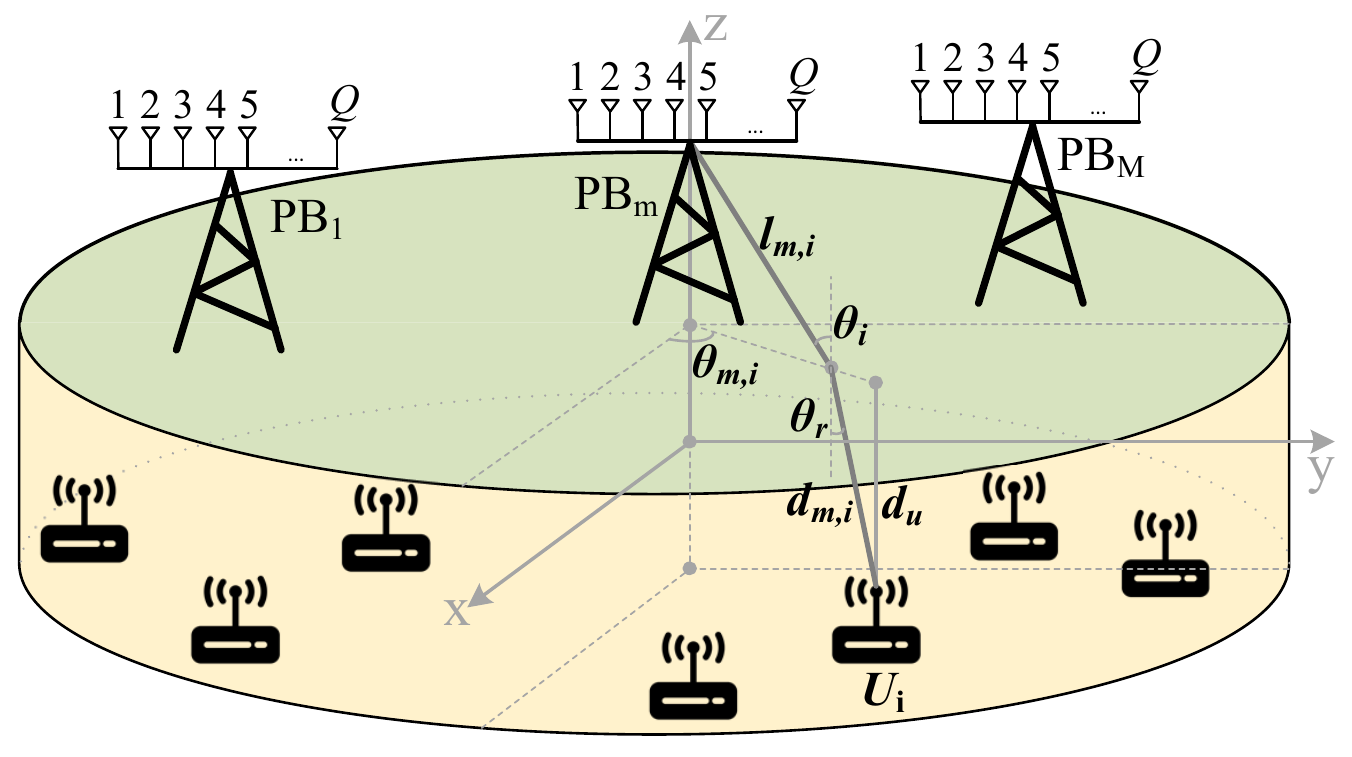}
    \caption{The distributed CSI-free system in a WPUSN: each ULA-equipped PB wirelessly charges nearby single-antenna underground EH devices.}
    \label{Systemfig}
\end{figure}

Consider the WPUSN depicted in Fig.~\ref{Systemfig} where a set $\mathcal{M}=\{\mathrm{PB}_{m}|m=1,2,\dots,M\}$ of PBs are deployed to wirelessly power a massive set $\mathcal{U}=\{U_{i}|i=1,2,\dots,N\}$ of single-antenna underground EH devices located nearby in a circle of radius $R$. Herein, we assume that all the devices are buried at the same depth $d_u$, and each PB is equipped with a uniform linear array (ULA) of $Q$ half-wavelength spaced antenna elements. 
\subsection{Channel Model}
\subsubsection{Small-Scale Fading Model} 
We assume quasi-static channels -- the channel coefficients remain constant over a transmission block and are independent and identically distributed (i.i.d) from block to block. Specifically, the channel is characterized by Rician fading, which allows modeling a wide variety of channels by adjusting the Rician factor $\kappa$. For instance, the channel envelope is Rayleigh distributed when $\kappa=0$; while there is a strong LOS channel when $\kappa \gtrsim 10$~\cite[Ch. 2]{Ricianfading}. 

The normalized channel vector between the ULA of $\mathrm{PB}_{m}$ and $U_{i}$ is expressed as~\cite[Ch. 5]{Ricianeq}
\begin{equation}
    \mathbf{h}_{m,i}\left(\theta_{m,i}\right) = \sqrt{\frac{\kappa}{1+\kappa}}\mathbf{h}_{m,i}^{\mathrm{los}}\left(\theta_{m,i}\right)+\sqrt{\frac{1}{1+\kappa}}\mathbf{h}^{\mathrm{nlos}},
\end{equation}
where $\mathbf{h}_{m,i}^{\mathrm{los}}\left(\theta_{m,i}\right)=e^{\mathbbm{i} \psi_{0}}\left[1, e^{\mathbbm{i} \phi_{1, m, i}}, e^{\mathbbm{i} \phi_{2, m, i}}, \ldots, e^{\mathbbm{i} \phi_{Q-1, m, i}}\right]^{\mathrm{T}}$ is the deterministic LOS component, while $\mathbf{h}_{\mathrm{nlos}}\sim\mathcal{CN}(\mathbf{0},\mathbf{R})$ accounts for the non-LOS (NLOS) channel under the scattering (Rayleigh) fading. More specifically, $\psi_{0}$ is an initial phase shift that can be ignored since it affects all antenna elements equally. Meanwhile, $\phi_{t, m, i}$, $t \in \{1, \ldots, Q-1\}$ represents the mean phase shift of the ($t+1$)th antenna element relative to the first antenna element as observed by $U_i$. That is~\cite{CSIfreeIoT} 
\begin{equation}
    \phi_{t, m, i} = -t \pi \sin(\theta_{m,i}),
\end{equation}
where $\theta_{m,i} \in [0, 2\pi]$ is the azimuth angle of $U_{i}$ concerning the transmitting ULA as illustrated in Fig.~\ref{Systemfig}, which depends on both PB's ULA orientation and device's location.

\subsubsection{Path Loss Model} 
The total path loss from $\mathrm{PB}_{m}$ to $U_{i}$, i.e., $ \delta_{m,i}$, comprises the above-ground air attenuation $L_{m,i}^{a}$, the refraction loss in the air-soil interface $L_{m,i}^{r}$, and the attenuation in underground soil $L_{m,i}^{u}$. Mathematically, the model is given by~\cite{LinLoRaWUSNs, AnnaWUSNs, Undergroundfield}  
\begin{align}
 \delta_{m,i} &= L_{m,i}^{a}L_{m,i}^{r}L_{m,i}^{u},\\
\label{airpathloss} L_{m,i}^{a}&=\left(\frac{4 \pi f}{c}\right)^2 \left(l_{m,i}\right)^\tau, \\
\label{Rpathloss} L_{m,i}^{r}&=\left(\frac{\sqrt{\left(\sqrt{\varepsilon'^{2}+\varepsilon''^{2}}+\varepsilon' \right) / 2}+1 }{4}\right)^2, \\
\label{soilpathloss} L_{m,i}^{u}&=\left(\frac{2 \beta d_{m,i}}{e^{-\alpha d_{m,i}}}\right)^{2},
\end{align}
where $f$ is the carrier frequency, $c$ denotes the speed of light in free space, $\tau$ is the path-loss exponent, while $l_{m,i}$ and $d_{m,i}$ are the air and underground soil propagation distances from $\mathrm{PB}_{m}$ to $U_{i}$, respectively. Since the permittivity of soil is much larger than air, most RF signal energy from the above-ground sink will be reflected back if the incident angle $\theta_i$ is large; therefore, we only consider the RF signal with a small incident angle, and the refracted angle $\theta_r$ is close to zero during the RF signal propagation from air to underground soil~\cite{Undergroundchannel, Undergroundfield}. Thus, we assume in this study that the propagation in the soil is vertical and $d_{m,i}$ equals the burial depth of devices $d_{u}$. Additionally, $\alpha$ and $\beta$ represent the attenuation constant and phase shifting constant, respectively, which are given as
\begin{align}
\alpha &= 2 \pi f \sqrt{\frac{\mu_{r} \mu_{0} \varepsilon' \varepsilon_{0}} {2}\left[\sqrt{1+\left(\frac{\varepsilon''}{\varepsilon'}\right)^{2}}-1 \right]}, \\
\label{beta}\beta &= 2 \pi f \sqrt{\frac{\mu_{r} \mu_{0} \varepsilon' \varepsilon_{0}} {2}\left[\sqrt{1+\left(\frac{\varepsilon''}{\varepsilon'}\right)^{2}}+1 \right]}.
\end{align}
Herein, $\mu_{r}$ is the soil’s relative permeability, $\mu_{0}$ is the free-space permeability, $\varepsilon_{0}$ is the free space permittivity, and $\varepsilon'$ and $\varepsilon''$ are the real and imaginary parts of the soil’s relative permittivity, respectively, i.e., $\varepsilon = \varepsilon' + j\varepsilon''$. Notice that  $\varepsilon$ can be calculated by the accurate mineralogy-based soil dielectric model~\cite{MBSDM}. For this, only three input parameters are required: the VWC, the operating frequency of the RF signals, and the percentage of clay in soil.

\subsection{Incident RF Power per Link} 
Let us focus on a single energy transfer link. Specifically, the signal $y_{m,i}$ received by $U_{i}$ when $\mathrm{PB}_{m}$ transmits $K \le Q$ energy symbols $\{s_{k}\}$ can be modeled by 
\begin{equation}
\label{signal}
    y_{m, i}=\sum_{k=1}^K \sqrt{\frac{p_{k}}{\delta_{m,i}}} \mathbf{v}_k^{\mathrm{T}} \mathbf{h}_{m,i} s_{k},
\end{equation}
where $\mathbf{v}_k = [v_{k}^{(1)}, v_{k}^{(2)},\ldots, v_{k}^{(Q)}] \in \mathbb{C}^Q$ is the normalized precoding vector associated with $s_{k}$, which depends on the adopted CSI-free scheme, while $p_{k}$ is the transmit power corresponding to each energy symbol and $\sum_{k=1}^{K}p_{k} = p_{m}, m \in [1, 2, \ldots, M]$. Here, we assume the same total transmit power for each PB, thus $p_m=p$, $\forall m$. Since the influence by noises is practically null and can be neglected in the EH process, we omitted its impact in~\eqref{signal}. The energy symbols $\{s_k\}$ are assumed to be i.i.d. unit-power and zero-mean random variables, i.e.,  $\mathbb{E}[|s_{k}|^2]=1$, $ \mathbb{E}[s_k]=0,$ and $\mathbb{E}[s_{k}^H s_{k^\prime}]=0~\forall k \neq k^\prime$~\cite{CSIfreeTcom,CSIfreeIoT,RAB, ClerckxCSIfree,CSIfreeTWC}. Therefore, the incident RF power (averaged over the signal waveform) at $U_{i}$,  which is triggered by $\mathrm{PB}_{m}$, is given by~\cite{RAB} 
\begin{align}
\label{pmi}
\xi_{m,i} &=\mathbb{E}_{s_k}\left[\left|y_{m, i}\right|^2\right] \nonumber \\
& \stackrel{(a)}{=} \mathbb{E}_{s_k}\!\! \!\left[\!\!\left(\sum_{k=1}^K \! \sqrt{\frac{p_{k}}{\delta_{m,i}}} \mathbf{v}_k^{\mathrm{T}} \mathbf{h}_{m,i} s_{k}\!\!\right)^{\mathrm{\! \! \! H}} \!\!\! \left(\sum_{k=1}^K \! \sqrt{\frac{p_{k}}{\delta_{m,i}}} \mathbf{v}_k^{\mathrm{T}} \mathbf{h}_{m,i} s_{k}\!\!\right)\!\!\right] \nonumber \\
& \stackrel{(b)}{=} \frac{1}{\delta_{m,i}} \sum_{k^{\prime}=1}^K \sum_{k^{\prime \prime}=1}^K \!\! \sqrt{p_{k^{\prime}} p_{k^{\prime \prime}}}\!\left(\mathbf{v}_{k^{\prime}}^{\mathrm{T}} \mathbf{h}_{m,i}\right)^{\!\mathrm{H}} \!\!\mathbf{v}_{k^{\prime \prime}}^{\mathrm{T}} \mathbf{h}_{m,i} \mathbb{E}\left[s_{k^{\prime}}^{\mathrm{H}} s_{k^{\prime \prime}}\right] \nonumber \\
& \stackrel{(c)}{=} \frac{1}{\delta_{m,i}} \sum_{k=1}^K p_{k}\left|\mathbf{v}_k^{\mathrm{T}} \mathbf{h}_{m,i}\right|^2,
\end{align}
where (a) comes from leveraging~\eqref{signal}, (b) follows after reorganizing terms, and (c) is obtained based on the assumption of i.i.d power-normalized signals.

In a typical quasi-static WET setup, the power harvested by $U_{i}$ converges to $E_{i}=g(\xi_{m,i})$, where $g$ is the function modeling the relationship between incident RF power and harvested power. In a linear EH model, the average harvested power is directly proportional to the average incident RF power, implying $g(\xi_{m,i})=\zeta \xi_{m,i}$, where $\zeta \in [0,1)$ denotes the energy conversion efficiency~\cite{LiuWPUSNs}. However, the EH process is non-linear in practice due to the non-linearities of the EH hardware; thus, non-linear EH models are intrinsically more accurate~\cite{ClerckxCSIfree, PracticalEH3, PracticalEH, PracticalEH2}. Nevertheless, since the harvested power benefits from an increased average incident RF power  (either under a linear or nonlinear EH model), we focus on the latter to reflect the performance of the proposed distributed CSI-free system in WPUSNs. 

\section{State-of-the-Art CSI-free Schemes} \label{CSIfreeSchemes}

\begin{figure}[t]
    \centering
    \includegraphics[width=3.45in]{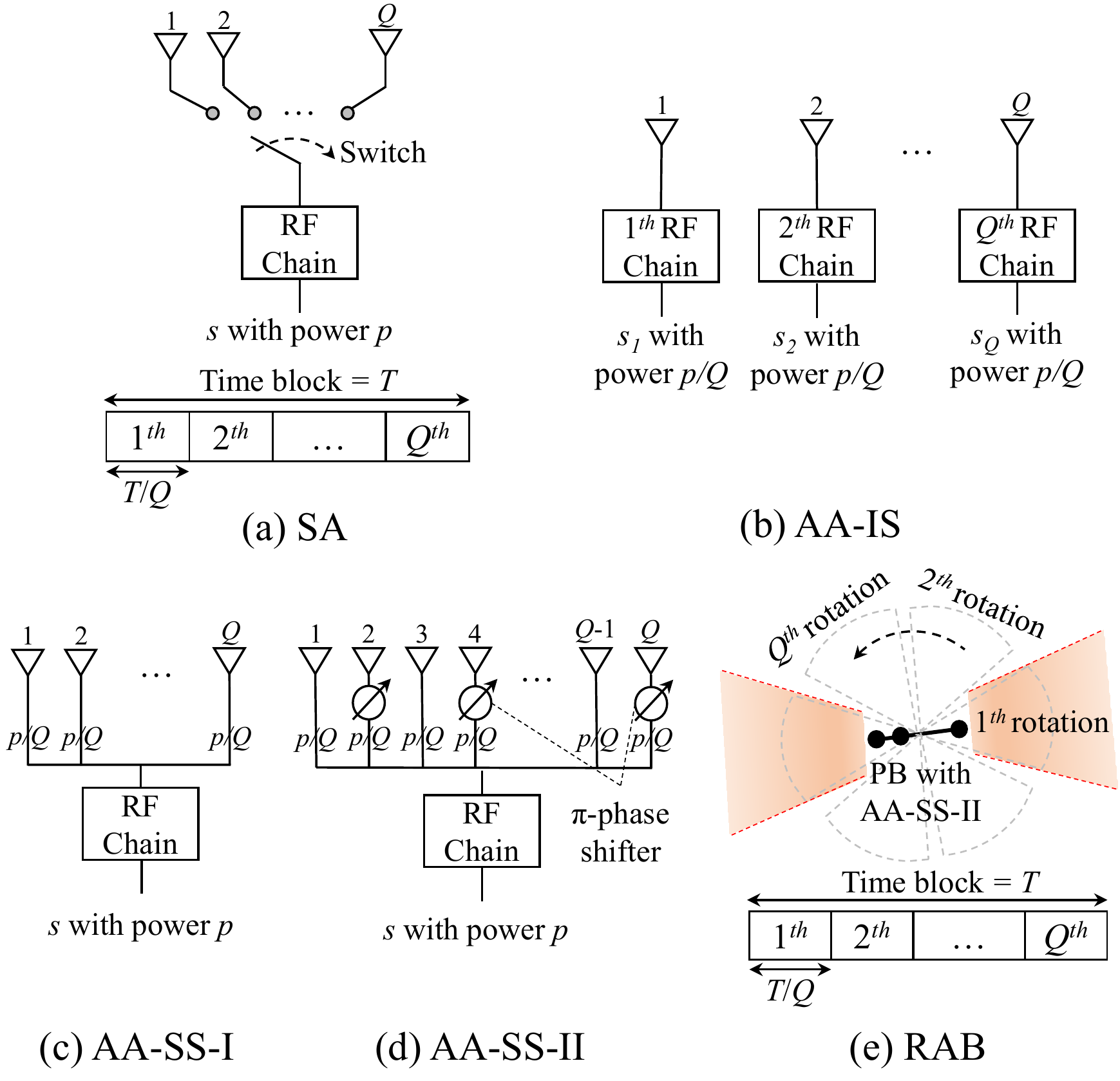}
    \caption{The operation diagram of (a) SA, (b) AA-IS, (c) AA-SS-I, (d) AA-SS-II, and (e) RAB.}
    \label{CSIfreeFig}
\end{figure}

Next, we briefly explain the state-of-the-art CSI-free schemes that will be adopted and comparatively analyzed in our proposed distributed CSI-free system. Figs.~\ref{CSIfreeFig} (a)-(e) illustrate the operations/implementations of SA, AA-IS, AA-SS-I, AA-SS-II, and RAB, respectively. We assume that each PB adopts the same CSI-free scheme.


\subsection{SA~\cite[Sec. III-A]{CSIfreeTcom}}
Under SA, the PBs utilize a switching mechanism to transmit a signal with the full power of the $q$th antenna at the $q$th duration such that $Q$ antennas are used during a normalized time block $T$, as depicted in~Fig.~\ref{CSIfreeFig}~(a). We assume equal time allocation for each antenna such that each subblock duration is equivalent to $T/Q$. Since only one antenna is active at the $q$th subblock duration, SA requires a single RF channel for its operation, implying $K=1$, $p_k=p$, and  $\mathbf{v}_{k}=[1]$. Note that the total incident RF energy should account for the sum of $Q$ subblocks.

\subsection{AA-IS~\cite[Sec. III-B]{CSIfreeIoT}}
Instead of transmitting a signal with one antenna at a time, the PB using AA-IS transmits signals independently generated across the antenna elements and with equal transmit power, thus $K=Q$ and $p_k = p/Q$. Therefore, $v_{k}^{(q)}=1$ for $k=q$, otherwise $v_{k}^{(q)}=0$. Different from SA, $Q$ RF chains are required to implement AA-IS since all the antenna elements are simultaneously active to transmit $Q$ independent RF signals, as highlighted in~Fig.~\ref{CSIfreeFig} (b). However, it is evidenced in~\cite{CSIfreeIoT} that SA has equal/similar WET performance to AA-IS under a linear/nonlinear EH model. Furthermore, the radiation patterns for both SA and AA-IS are omnidirectional.

\subsection{AA-SS}
In AA-SS, the same signal is transmitted through all antenna elements with equal power, i.e., $K=1$ and $p_k = p/Q$. There are two configurations for AA-SS:
\begin{enumerate}[itemindent=1em]
    \item AA-SS-I~\cite[Sec. III-A]{CSIfreeTcom}, where the precoding vector $\mathbf{v}_{k}=\mathbf{1}$, or simply no precoder, is applied to attain an energy beam toward the ULA's boresight directions, as displayed in ~Fig.~\ref{CSIfreeFig}~(c).
    
    \item AA-SS-II~\cite[Sec. IV-B]{CSIfreeIoT}, where the precoding vector is set as $v_{k}^{(q)}=e^{\mathrm{mod}(q-1,2)\pi \mathbbm{i}}$ to attain wider energy beams, which are offset $90^\circ$ from ULA's boresight directions. This can be realized with an analog implementation with a number of $\lfloor Q/2\rfloor$ $\pi$-phase shifters as shown in~Fig.~\ref{CSIfreeFig}~(d).
\end{enumerate}

The gains of both AA-SS schemes are strongly associated with the EH devices' positions and the orientation of the PB's antenna array; therefore, they are preferable when charging EH devices clustered in specific boresight directions. Consequently, the performance of both schemes can be further improved by properly rotating the PB antenna array or optimizing PB's deployments based on the positioning information of the devices.
\subsection{RAB~\cite{RAB}}
A servo motor is equipped in each PB to continuously rotate its antenna array while adopting the AA-SS-II scheme, which allows improving the charging coverage probability. By taking advantage of the symmetry of the ULA’s radiation patterns, we consider $Q$ angular rotations to cover the angular domains $[-\pi/2, \pi/2]$. Indeed, a servo motor can rotate the antenna array at specific angles and realize $Q$ equally spaced steps using the pulse with modulation (PWM) technique, that can provide sufficiently smooth performance. Note that the $q$th rotation step is conducted during the $q$th subblock duration, where each duration is $T/Q$, as exhibited in~Fig.~\ref{CSIfreeFig}~(e). Consequently, the incident average RF power gain at $U_{i}$ due to the transmissions of $PB_{m}$ is given by
\begin{align}
\label{RABenergy}
\xi_{m,i}^{RAB}= &\frac{1}{Q} \sum_{q=1}^Q  \sum_{k=1}^K \frac{p_{k}}{\delta_{m,i}} \left|\mathbf{v}_k^{\mathrm{T}}\mathbf{h}_{m,i}\left(\theta_{m,i}+\frac{q \pi}{Q}\right)\right|^2,
\end{align}
where $\theta_{m,i}$ is the initial azimuth angle prior to any rotation. The resulting radiation pattern is quasi-omnidirectional. Note that the operation of RAB requires at least $Q=2$ antenna elements; otherwise, it is equivalent to AA-SS-II.

\section{On the Multi-PB Deployment} \label{DPCS}
The appropriate deployment of multiple PBs is undoubtedly appealing for the CSI-free schemes to increase the low WET efficiency and eliminate the heterogeneous energy distribution over the charging area~\cite{DistributedWET1, DistributedWET2}, especially in the high-attenuation WPUSN scenarios. Motivated by this, we advocate the integration of CSI-free schemes and multi-PB deployments, implying a distributed CSI-free system. 

Ignore the energy contributed by coexisting/neighboring networks operating in the same spectrum and assume that all the PBs generate and transmit independent signals, the incident RF energy available at $U_{i}$ in the distributed CSI-free system is given by~\cite{DistributedWET1}   
\begin{equation}
    G_{i} = \sum_{m=1}^{M}\xi_{m,i}.
\end{equation}

Herein, having more PBs increases the total power budget and improves the WET coverage probability. A significant performance gain shall be attained in relatively static WET scenarios by optimizing the deployment of PBs. As evidenced in different scenarios~\cite{DistributedWET1, DistributedWET2, DistributedWET3, DistributedWET4, DistributedWET5, DistributedWET6, EFFCscheme}, the terrestrial WET coverage probability benefits from an efficient PBs' deployment strategy. However, the performance of such deployment approaches is not clear in the underground distributed CSI-free system. Motivated by this, we consider the well-known K-Means clustering algorithm based on prior positioning information of devices~\cite{Kmeans} and the position-agnostic EFFC approach~\cite[Sec. III-A]{EFFCscheme} for such a system. 

\subsection{K-Means}
K-Means clustering is an iterative data-partitioning algorithm~\cite{Kmeans}. The basic concept is that $N$ observations are assigned to exact one of $M$ clusters defined by centroids such that the sum of points-to-centroid distances is minimized. In this study, an observation constitutes an underground EH device, while each PB is a centroid. Recall that we focus on practical scenarios where the devices are buried at the same depth to monitor the same type of underground entities such as crop roots and gas/oil pipeline. Therefore, the K-Means approach requires only the number of PBs $M$ and the above-ground projection of each underground EH device to determine the optimal PBs' deployment. It is worth noting that although the underground devices' positions can be obtained from the prior deployment information or be collected via geophysical surveys, it might remain challenging as the position accuracy may deteriorate due to the deformation or movement of the soil, and more importantly, additional energy consumed by position information acquisition would impair massive WPUSNs scenarios.


Let $\mathbf{u_n}, \mathbf{b_m} \in \mathbb{R}^{2\times1}$ denote coordinate vectors for the locations of $U_n$ and $\mathrm{PB}_{m}$, respectively. Algorithm \ref{KMeansalg} illustrates the procedure for determining the optimal positions of PBs by adopting the K-Means approach. First, the positions of $M$ PBs are randomly selected in line~\ref{KS1}. Then, each device is assigned to its closest PB according to the Euclidean distance in line~\ref{KS2}. Next, line~\ref{KS3} is for updating the positions of the PBs by computing the average position of the devices per cluster. Finally, lines~\ref{KS2}, and~\ref{KS3} are repeated until no device changes its associated PB, after which the optimal PBs' positions are obtained.

\begin{algorithm}[t]
    \begin{footnotesize}
	\caption{K-Means for determining the PBs' positions}
	\label{KMeansalg}
     \renewcommand{\algorithmicrequire}{\textbf{Input:}}  
    \renewcommand{\algorithmicensure}{\textbf{Output:}} 
	\begin{algorithmic}[1]
	\REQUIRE $\mathbf{u_1},\ldots,\mathbf{u_N}, M$
    \STATE \label{KS1} Select randomly $M$ coordinates of above-ground PBs' positions $\mathbf{b_1},\ldots,\mathbf{b_M}$
    \REPEAT
    \STATE \label{KS2} Set $a_{n,m}\!=\!1$ if $m\!=\!\underset{j=1,\ldots,M}{\mathrm{argmin}} \|\mathbf{u_n}\!-\!\mathbf{b_j}\|$, otherwise $a_{n,m}\!=\!0$, $\forall n, m$
        \STATE \label{KS3} Set $\mathbf{b_m} = \frac{\sum_{n=1}^N a_{n,m} \mathbf{u_n}}
            {\sum_{n=1}^N a_{n,m}}$, $\forall m$
    \UNTIL{Convergence}
    \ENSURE $\mathbf{b_1},\ldots,\mathbf{b_M}$
	\end{algorithmic}  
     \end{footnotesize}
\end{algorithm}
\subsection{EFFC}
Recently, Rosabal~\textit{et al.}~\cite{EFFCscheme} proposed a low-complexity PBs deployment approach, i.e., EFFC, to maximize the charging coverage area in a system where the PBs are not aware of the devices' locations. Specifically, there are two possible configurations in EFFC: 
\begin{enumerate}
    \item EFFC-I, where the PBs are placed equally-far-from the circle center at a distance $r = \|b_m\|,~0 \leq r \leq R$, with the angle $\varphi=2\pi/M$ between any two adjacent PBs;
    \item EFFC-II, where one PB is located at the center and the remaining PBs are deployed at a distance $r$ from the circle center with $\varphi=2\pi/(M-1)$. 
    \end{enumerate}

Only four input parameters are required in EFFC to determine the PBs position: i) the number of PBs $M$, ii) the path-loss exponent $\tau$, iii) the radius of the service area $R$, and iv) the step size of the algorithm $\triangle r$. Since EFFC is conceptually simple and does not require prior information about the devices' positions, it may facilitate practical deployments of distributed CSI-free systems in WPUSNs, especially for powering a massive number of underground EH devices. 

Algorithm~\ref{EFFCalg} details how EFFC determines the suitable configuration and the optimal PBs' positions for our considered scenario. In lines~\ref{Ec} and~\ref{Ee1}, we set $E_{c}$ and $E_{e1}$ as the estimates of the RF energy that would be available at a sensor hypothetically located on the center and edge of the area, respectively, in EFFC-I. For EFFC-II, $E_{x}$ given in~line~\ref{Ex} denotes the estimate of the RF energy that would be available at a sensor that is equidistant to the center and two adjacent PBs at a distance $x = r/(2\cos(\frac{\pi}{M-1}))$, while $E_{e2}$ given in~line~\ref{Ee2} represents the estimate of the RF energy that would be available at a sensor on the edge. In line~\ref{ES1}, we adopt the EFFC-I to determine the positions of the PBs if the contribution of EFFC-I is higher than that of EFFC-II under $r$. On the other hand, when EFFC-II outperforms EFFC-I under $r$, the PBs' deployment is based on  EFFC-II, as in line~\ref{ES2}. We repeat lines~\ref{Ec}-\ref{radd} until $r$ surpasses $R$ to then attain the optimal deployment of the PBs.

\begin{algorithm}[t]
\begin{footnotesize}
\caption{EFFC for determining the PBs' positions}
	\label{EFFCalg}
     \renewcommand{\algorithmicrequire}{\textbf{Input:}}  
    \renewcommand{\algorithmicensure}{\textbf{Output:}} 
	\begin{algorithmic}[1]
	\REQUIRE $M, \tau, R, \triangle r$ 
    \STATE Set $r^*=0, r=\triangle r, \varphi=0$, $E^*=MR^{-\tau}, x = \frac{r}{2\cos\left(\frac{\pi}{M-1}\right)}$
    \REPEAT
    \STATE \label{Ec} $E_c=Mr^{-\tau}$
    \STATE \label{Ee1} $E_{e1}=\sum_{m=1}^{M}\left[r^2+R^2\!-\!2rR\cos\left(\frac{2\pi}{M}(m-\frac{3}{2})\right)\right]^{-\frac{\tau}{2}}$
    \STATE \label{Ex} $E_x=x^{-\tau}+\sum_{m=1}^{M-1}\left[x^2\!+r^2-2xr\cos\left(\frac{2\pi}{M-1}\left(m-\frac{3}{2}\right)\right)\right]^{-\frac{\tau}{2}}$
    \STATE \label{Ee2} $E_{e2}=R^{-\tau}+\sum_{m=1}^{M-1}\left[R^2\!+r^2-2Rr\cos\left(\frac{2\pi}{M-1}\left(m-\frac{3}{2}\right)\right)\right]^{-\frac{\tau}{2}}$
        \IF {$\min(E_c,E_{e1})>\min(E_x,E_{e2})$} \label{EIFbegin}
           \STATE \label{ES1} Set $E^*=\min(E_c,E_{e1})$, $\varphi=\frac{2\pi}{M}$, $r^*=r$,  \qquad\qquad \qquad \qquad $\mathbf{b_m} = [r^*\cos(\varphi(m-1)), r^*\sin(\varphi(m-1))], \forall m$ \qquad \qquad \qquad if $E^* < \min(E_c,E_{e1})$
        \ELSE
                
           \STATE \label{ES2} Set $E^*=\min(E_x,E_{e2})$, $\varphi=\frac{2\pi}{M-1}$, $r^*=r$,  \qquad \qquad \qquad   $\mathbf{b_m}\!=\!\left\{ \begin{aligned}
&[0, 0], m=1\\ &[r^*\cos(\varphi(m-2)), r^*\sin(\varphi(m-2))], \text{otherwise} \end{aligned} \right. \forall m$ \qquad if $E^* < \min(E_x,E_{e2})$
        \ENDIF \label{EIFend}
        \STATE $r:=r+\triangle r$ \label{radd}
    \UNTIL{$r\geq R$}
    \ENSURE $\mathbf{b_1},\ldots,\mathbf{b_M}$
	\end{algorithmic}  
 \end{footnotesize}
\end{algorithm}

\section{Power Budget Model} \label{PowerBudget}
In this section, we consider a power budget for each PB to execute the complete WET operations. We model the relationship between the budgeted power and the total transmit power for the proposed distributed CSI-free systems. 

\subsection{Ideal System}
In the ideal system, the budgeted power $P_{budget}$ allocated for each PB is entirely converted into the total transmit power without any loss, i.e., $p = P_{budget}$.

\subsection{Practical System} \label{Secpracticalsys}
Due to the power consumed by the transmitter power amplifier, circuitry, and operations, the budgeted power is not completely available for transmission in a practical system. Herein, we consider the impact of the circuitry and CSI-free base-band operations on the power consumption. Note that the power consumed for the base-band signal processing in AA-IS is the greatest in comparison with SA, AA-SS-I, AA-SS-II, and RAB. This is because $Q$ RF chains are needed to implement AA-IS, while only one RF chain is needed to implement the other CSI-free schemes. Compared with the power consumption in RF chains, the power consumed by the switch (in the case of SA) and phase shifters~\footnote{The $\pi$-phase shifters in AA-SS-II and RAB can be implemented passively with a resistor-capacitor phase-shift network at microwave frequencies, or simply with a half-wavelength delay line at higher frequencies, implying negligible power consumption.} (in the case of AA-SS-II and RAB) is negligible and we do not consider it in this study~\cite{Switchpower,Phasepower1,Phasepower2}. Therefore, the total transmit power of each PB in the practical distributed CSI-free system is given by~\cite{TxPower} 
\begin{equation}
    p = \eta (P_{budget}-Q^\varsigma P_{rf}-P_c),
    \label{p_pra}
\end{equation}
where $\eta$ is the amplifier efficiency, $P_{rf}$ is the power consumed by the base-band processing per RF chain, $P_c$ is a fixed power consumption considering the remaining circuitry, and $\varsigma=1$ for AA-IS, and $\varsigma=0$ for the other CSI-free schemes. Here, $P_c$ and $P_{rf}$ are assumed to be constant without loss of generality.

\begin{figure} [t]
    \centering
    \includegraphics[width=3.45in]{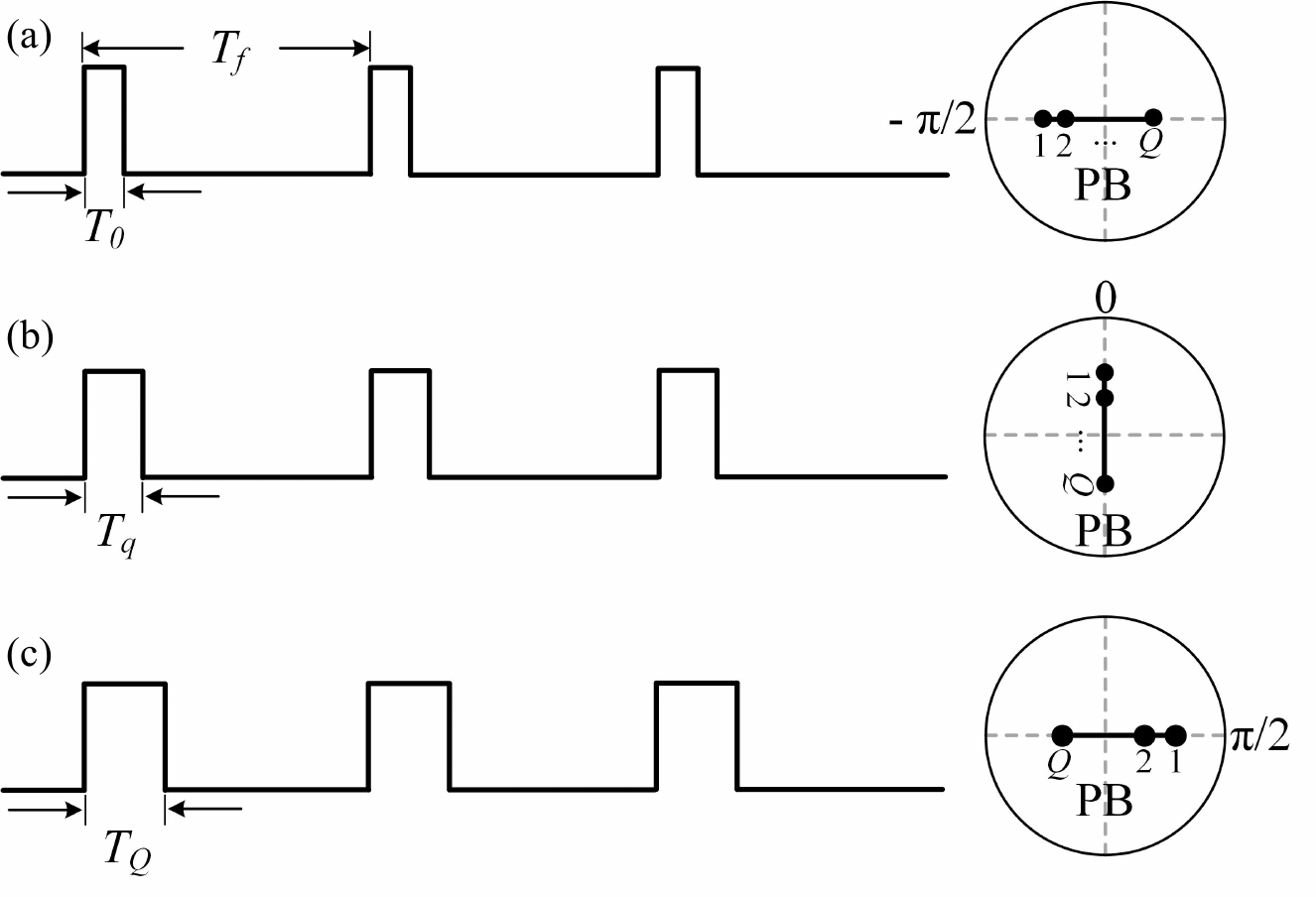}
    \caption{The operations of the servo motor by using the PWM principle given $Q=2$. Specifically, (a) a $T_0$ pulse positions the shaft at $-\pi/2$ (top), then (b) a pulse with $T_q = T_0+\frac{T_Q-T_0}{2}$ rotates the ULA by $\pi/2$ (middle), and finally (c) a $T_Q$ pulse triggers the ULA's $Q$th rotation step under $T_{f}$ (bottom).}
    \label{Motoroperation}
\end{figure}

In addition, compared with the SA, AA-IS, AA-SS-I, and AA-SS-II schemes, the extra power consumed by the servo motor operations $P_{motor}$ should be carefully considered in the RAB scheme. By using the PWM principle, the servo motor can rotate an antenna array with $Q$ equally spaced steps in the angular domains $[-\pi/2, \pi/2]$. More specifically, the servo motor controls the rotation angle by adjusting the width of an electrical pulse applied to the control wire. As depicted in Fig.~\ref{Motoroperation}, a servo motor excited with a given pulse width $T_q$ triggers the ULA's $q$th rotation step, which we set as $q\pi/Q, q \in [1, 2, \ldots, Q]$. The pulse detection frequency is 1/$T_f$, where $T_f$ is the duty cycle. Since one rotation step of the servo motor requires at least a complete duty cycle, the subblock duration as described in~Fig.~\ref{CSIfreeFig}~(e) should be longer than $T_f$, implying $T/Q \geq T_f$. Note that the servo motor should apply a $T_0$ pulse for locating the beginning angle before the rotation operations. Therefore, by considering practical servo motors, the total transmit power of each PB under RAB can be modeled as
\begin{align}
    p^{RAB} &= \eta (P_{budget}-P_{rf}-P_{motor}), \label{RABEnergy1} \\
    P_{motor} &= \frac{T_0+\sum_{q=1}^{Q}T_{q}}{T_{f}}V_{supply}I_{work},\label{RABEnergy2} \\
    T_{q} &= T_0+q\frac{(T_Q-T_0)}{Q} \label{RABEnerg3},
\end{align}
where $T_0$ and $T_Q$ are the pulse width for the shaft at $-\pi/2$ and $\pi/2$, respectively. $V_{supply}$ is the supply voltage of the servo motor, and $I_{work}$ is the working current during rotation.

\section{Numerical Results} \label{NumRes}
This section presents our numerical results on the performance of the proposed distributed CSI-free system in WPUSNs by considering different network configurations and underground parameters. We consider an EH threshold $\gamma$ such that the devices can harvest energy only when the incident RF energy exceeds $\gamma$~\cite{EHthreshold}. Additionally, we compare the discussed CSI-free schemes and evaluate their performance compared to a full-CSI strategy. The latter scheme optimizes a CSI-based precoder designed to charge the underground EH devices with maximum fairness. This optimization problem can be formulated as a semi-definite programming (SDP) problem~\cite{OnelRician}, which in turn can be efficiently solved using CVX tools~\cite{cvx}. To further tilt the scale in favor of the CSI-free schemes, we analyze the performance of the full-CSI strategy without considering the power consumed for both the CSI acquisition and the SDP-based solution, which is consistent with the setup in~\cite{OnelRician, RAB, OnelWETReview}\footnote{Notice that assessing the power consumed in the CSI acquisition phase is not straightforward as it may depend on several system parameters that are not considered in this work, e.g., pilot transmit power, number of pilot symbols, and multi-user pilot scheduling. In addition, the impact of imperfect CSI should be considered. Therefore, we considered an ideally-trained full-CSI strategy, thus somewhat tilting the scale in its favor.}. Moreover, we assume the full-CSI strategy is implemented in fully-digital ULAs. Thus, the total transmit power of each PB for the full-CSI strategy is the same as the AA-IS, i.e., $\varsigma=1$ in~\eqref{p_pra}.

\begin{table}[t]
\caption{Simulation Parameters for CSI-free WPUSNs system} 
\centering
\begin{tabular}{m{0.28\textwidth}<{\raggedright} m{0.15\textwidth}<{\centering}}%
\toprule
\textbf{Parameters}                & \textbf{Values}                    \\  \hline
Deployment Radius ($R$)            & 5~m                                \\    
Total number of nodes ($N$)        & 64                                 \\
Nodes' deployment                  & uniform and random                 \\
Burial depth ($d_{u}$)             & 35~cm                              \\
Number of PBs ($M$)                & 1                                  \\
Number of antennas per PB ($Q$)    & 4                                  \\
VWC ($m_v$)                        & 15\%                               \\
Clay ($m_c$)                       & 38\%                               \\
Carrier center frequency ($f$)     & 433~MHz                            \\
Rician factor ($\kappa$)           & 10                                 \\      
Path-loss exponent ($\tau$)        & 2                                  \\ 
Power Budget ($P_{budget}$)        & 10~W                               \\
Amplifier efficiency ($\eta$)      & 38\%                               \\
Circuit power ($P_c$)              & 0.1~W                              \\
RF base-band consumption power ($P_{rf}$) & 0.06~W                      \\
Motor's duty cycle ($T_{f}$)       & 20~ms                              \\
Motor's voltage ($V_{supply}$)     & 5~V                                \\
Motor's current ($I_{work}$)       & 250~mA                             \\
EH threshold  ($\gamma$)           & -22~dBm                            \\                       
\bottomrule
\label{tab1}
\end{tabular}
\end{table}

To evidence the practical system's performance, we select a real-life center-pivot irrigation farm as the study scenario, where the scenario is assumed to be spacious and without obstacles, and $N=64$ underground EH devices are uniformly and randomly distributed in a 5~m-radius circular area. Note that the network density is close to 1 device/m$^2$, which conforms with the massive WET scenarios~\cite{mMTCscale}. Herein, the \textit{in-situ} soil characteristics obtained from~\cite{Undergroundfield}, e.g., clay percentage of soil, are applied to estimate the attenuation in soil. The antennas equipped in the PBs are deployed at 1.5 m above the ground surface, thus, channels may experience strong LOS propagation. Specifically, we set $\kappa=10$ and consider the path-loss exponent $\tau=2$~\cite{RAB}. The system's frequency band is set at 433 MHz, which is preferable in underground wireless communications~\cite{LinLoRaWUSNs}. We assume a power budget of 10~W per PB, while $\eta=38\%$, $P_c=0.1$~W, and $P_{rf}=0.06$~W~\cite{TxPower, Prf}. Unless stated otherwise, $M=1$ PB equipped with $Q=4$ antennas is located in the center of the service area, while VWC and burial depth are set as $m_{v}=15\%$ and $d_u=0.35$~m, respectively. For the RAB implementation, we consider a practical servo motor, i.e., Micro SG90, with the rotation range of $[-\pi/2, \pi/2]$. Here, a $T_0=1$~ms pulse positions the shaft at $-\pi/2$ while the shaft at $\pi/2$ is realized by a $T_Q=2$~ms pulse with $T_{f}=20$~ms~\cite{ServoMotor}. Here, the supply voltage and rotating current of Micro SG90 are 5~V and 250~mA, respectively. The simulation parameters are listed in Table~\ref{tab1}.


\subsection{On the Charging Coverage of the CSI-free Schemes per PB}
\begin{figure}[t]
    \centering
    \includegraphics[width=3.45in]{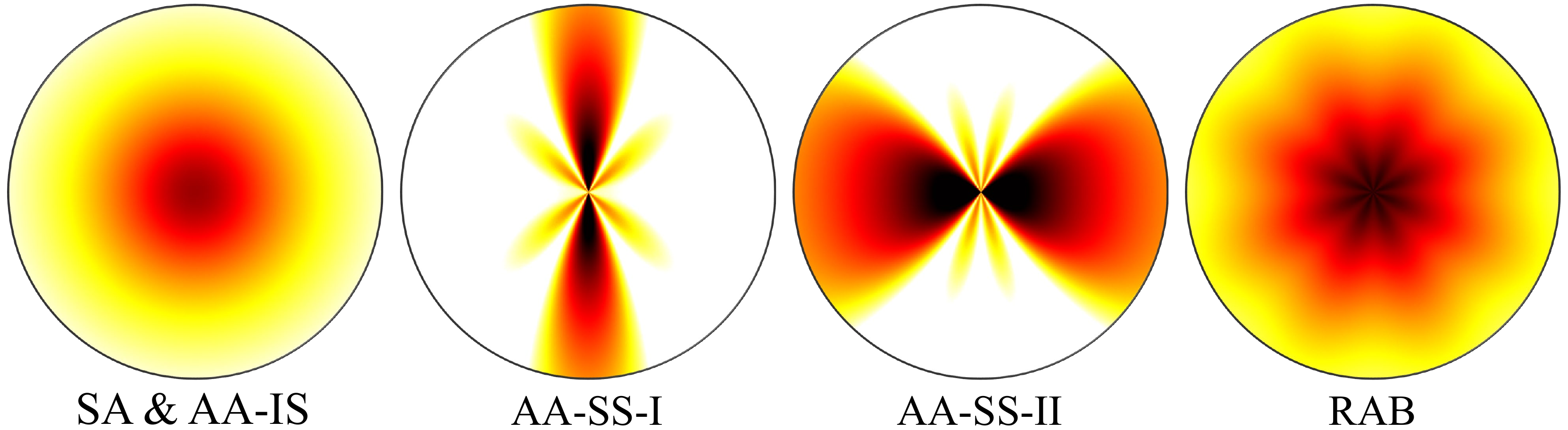}
    \caption{Heatmap of the average RF energy availability in dBm under the discussed CSI-free schemes and considering the ideal system. Here, a single PB equipped with $Q=4$ antennas is deployed in the center of the 5~m-radius area, and the devices are buried at the depth of 35~cm with $m_v=15\%$.}
    \label{Heatmap}
\end{figure}

\begin{figure*}[t]
    \centering
    \subfigure{\includegraphics[height=2.4in]{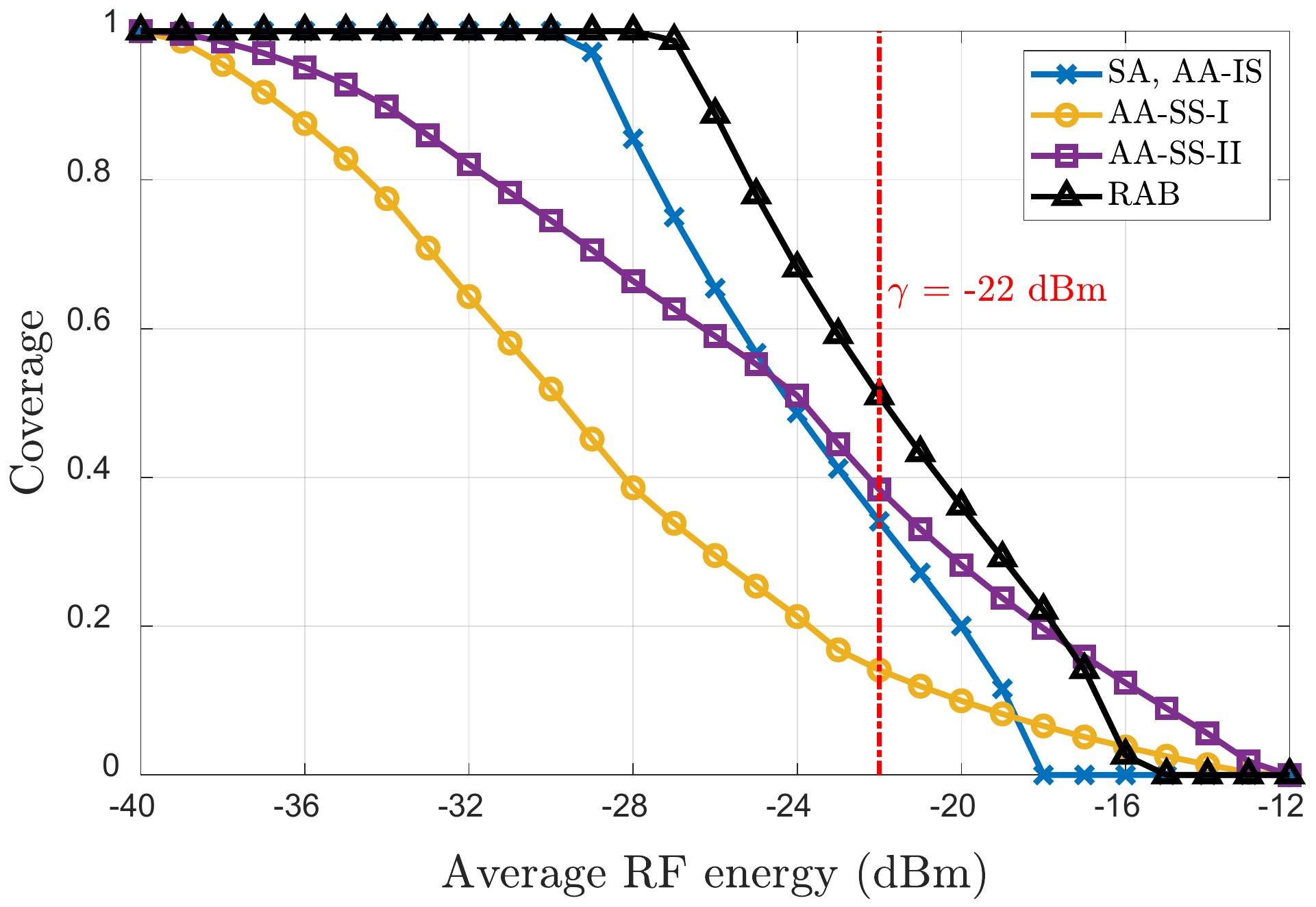}}
    \subfigure{\includegraphics[height=2.4in]{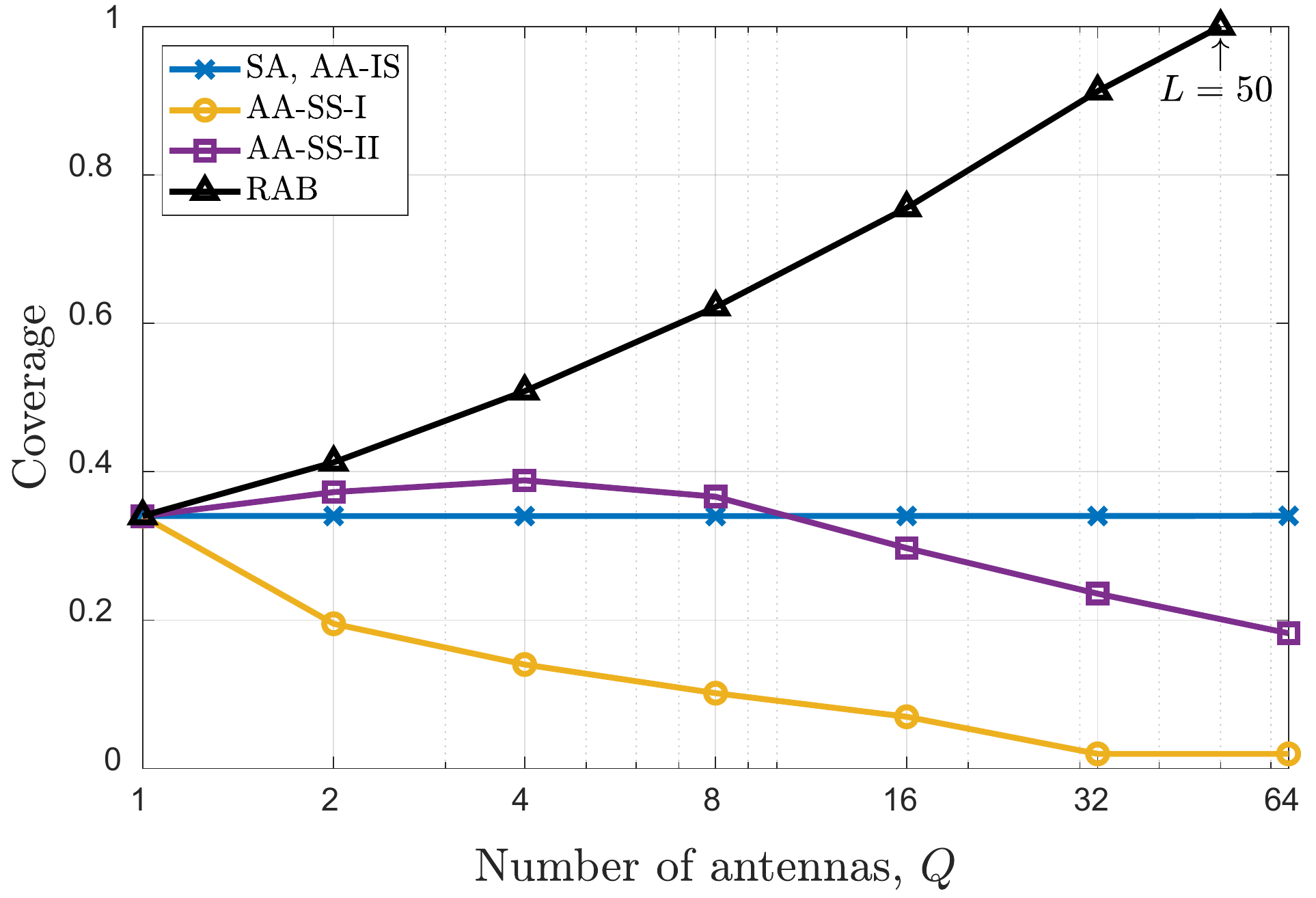}}
    \caption{(a) Area coverage for different average RF energy requirements with the number of antennas $Q=4$ (left), and (b) area coverage as a function of the number of antennas $Q$ given an EH threshold $\gamma=-22$~dBm (right). Here, we consider the discussed CSI-free schemes under the ideal system, where a single PB is deployed in the center of the 5~m-radius area, and the devices are buried at the depth of 35~cm with $m_v=15\%$. The red dash-dotted line depicts the EH threshold of $\gamma=-22$~dBm.}
    \label{CoverageRes}
\end{figure*}

Figs.~\ref{Heatmap} and~\ref{CoverageRes} show the average RF energy availability of different CSI-free schemes in a 5-m radius area served by a single PB located at the center, for the ideal system. More specifically, Fig.~\ref{Heatmap} exhibits the heatmap of the average RF energy availability in the service area with the number of antennas $Q=4$. Note that SA, AA-IS, and RAB provide a uniform performance along the area, i.e., omnidirectional/quasi-omnidirectional radiation patterns, while AA-SS-I and AA-SS-II favor the ULA boresight and 90$^\circ$ offset from ULA boresight directions, respectively. Therefore, both AA-SS schemes are preferable in certain deployments where the underground EH devices are clustered in specific spatial directions. 

Meanwhile, the statistics information on the average RF energy availability in the charging area for Fig.~\ref{Heatmap} is illustrated in Fig.~\ref{CoverageRes}~(a). It can be seen that AA-SS-I allows covering up to 14\% of the area with the EH threshold of $-22$~dBm. Such coverage can be increased to 34\% when operating with SA and AA-IS, and it further achieves 38\% and 51\% under AA-SS-II and RAB, respectively. Furthermore, under a looser EH threshold (e.g., $\gamma \leq -25$~dBm), SA, AA-IS and RAB featuring the omnidirectional/quasi-omnidirectional radiations pattern perform better in terms of coverage probability, compared to AA-SS-I and AA-SS-II.

Fig.~\ref{CoverageRes}~(b) depicts the charging coverage probability as a function of the number of antennas $Q$ given an EH threshold $\gamma=-22$~dBm. Observe that as the number of antennas $Q$ increases, the performance of AA-SS-I deteriorates, while the coverage probability of AA-SS-II increases with $Q$ only for $Q \leq 4$, and then diminishes with the increased $Q$. The reason for this is that a higher number of antennas leads to narrower beams, thus hampering a wide energy coverage. Meanwhile, the coverage for the AA-IS and SA schemes remains approximately the same as they implement omnidirectional radiation. Notably, RAB outperforms the other CSI-free schemes as its coverage probability increases with $Q$. Note that under $T_{f}=20$~ms, the allowed maximum number of antennas per PB in RAB is $Q=50$ by considering a normalized time block, i.e., $T=1$~s. 

It is worth noting that here we are considering only an ideal system, while the energy consumed by the servo motor components in RAB, especially as $Q$ increases, may potentially eliminate the aforementioned performance gains. Therefore, it is challenging for a single PB to guarantee the harvesting levels above $\gamma=-22$~dBm. Indeed, such a configuration is impractical for realizing the massive charging of devices in underground conditions, especially given higher VWC and larger burial depth. Consequently, we discuss the performance of distributed CSI-free systems in WPUSNs in~Sections~\ref{IdealSec}, \ref{SecPracticalRes}, and~\ref{Secunderres}.  


\subsection{On the Impact of Rician Factor $\kappa$}
\begin{figure}
    \centering
    \includegraphics[width=3.45in]{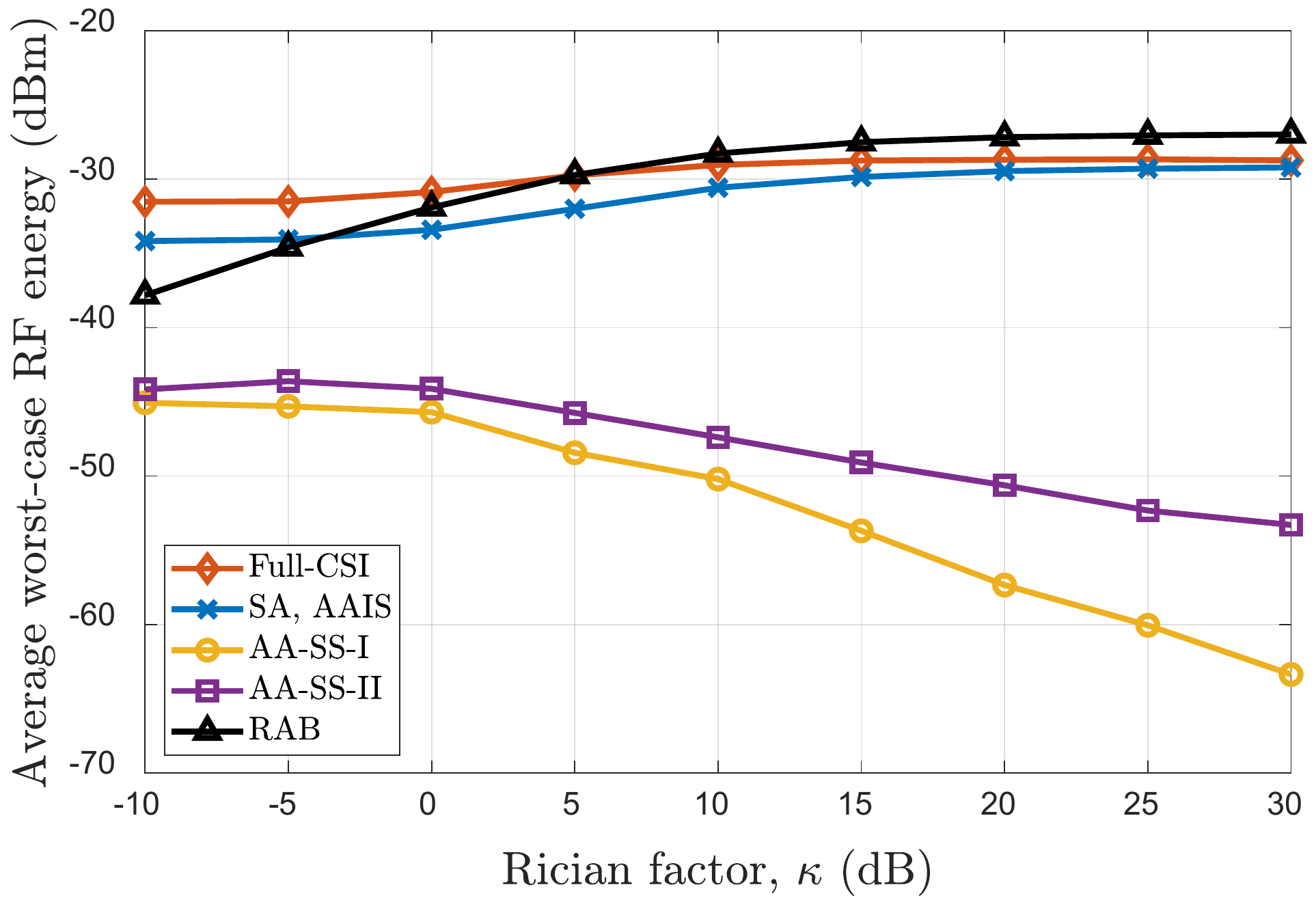}
    \caption{Average worst-case RF energy available as a function of Rician factor $\kappa$ when operating the full-CSI strategy and the discussed CSI-free schemes under the ideal system. Here, a single PB equipped with $Q=4$ antennas is deployed in the center of the 5~m-radius area, and the devices are buried at the depth of 35~cm with $m_v=15\%$.}
    \label{Kfactorfig}
\end{figure}

Fig.~\ref{Kfactorfig} illustrates the average worst-case RF energy when utilizing the discussed CSI-free schemes and the full-CSI strategy in a single-PB ideal system as a function of the Rician factor $\kappa$. Observe that the average worst-case RF energy under AA-SS-I and AA-SS-II deteriorates as the Rician factor increases. This is due to the fact that a larger Rician factor leads to more directional, but spatial-specific, beamforming, thus, a smaller coverage area. On the other hand, the performance of RAB, SA, AAIS, and the full-CSI strategy improves with $\kappa$. Specifically, the full-CSI strategy (without considering the CSI training energy costs) outperforms  the CSI-free schemes when $\kappa \leq 0$~dB, while RAB leads to the highest average RF energy when $\kappa \geq 5$~dB. These observations reveal that RAB can efficiently charge massive underground EH devices under strong LOS conditions. On the other hand, the full-CSI strategy may be strictly required for serving the underground EH devices when considering the harsh channels, i.e., $\kappa\leq 0$~dB, despite it necessitates significant energy consumption for the CSI acquisition and precoding design via SDP. 

\subsection{On the Ideal CSI-free Distributed System} \label{IdealSec}
\begin{figure*}[t]
	\centering
	\subfigure[$M=1$]{
    \includegraphics[width=2.3in]{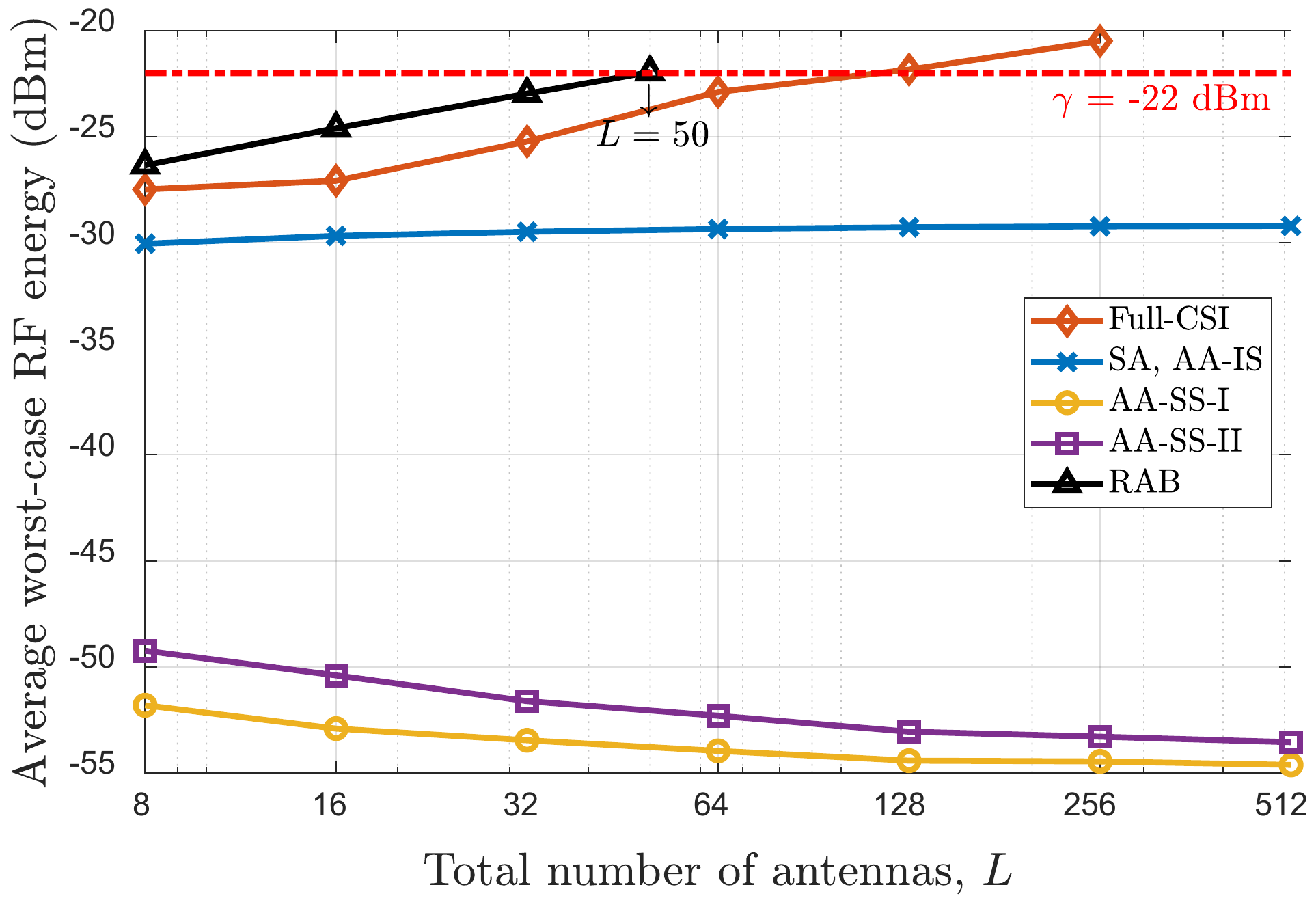}}
    \subfigure[$M=4$]{
    \includegraphics[width=2.3in]{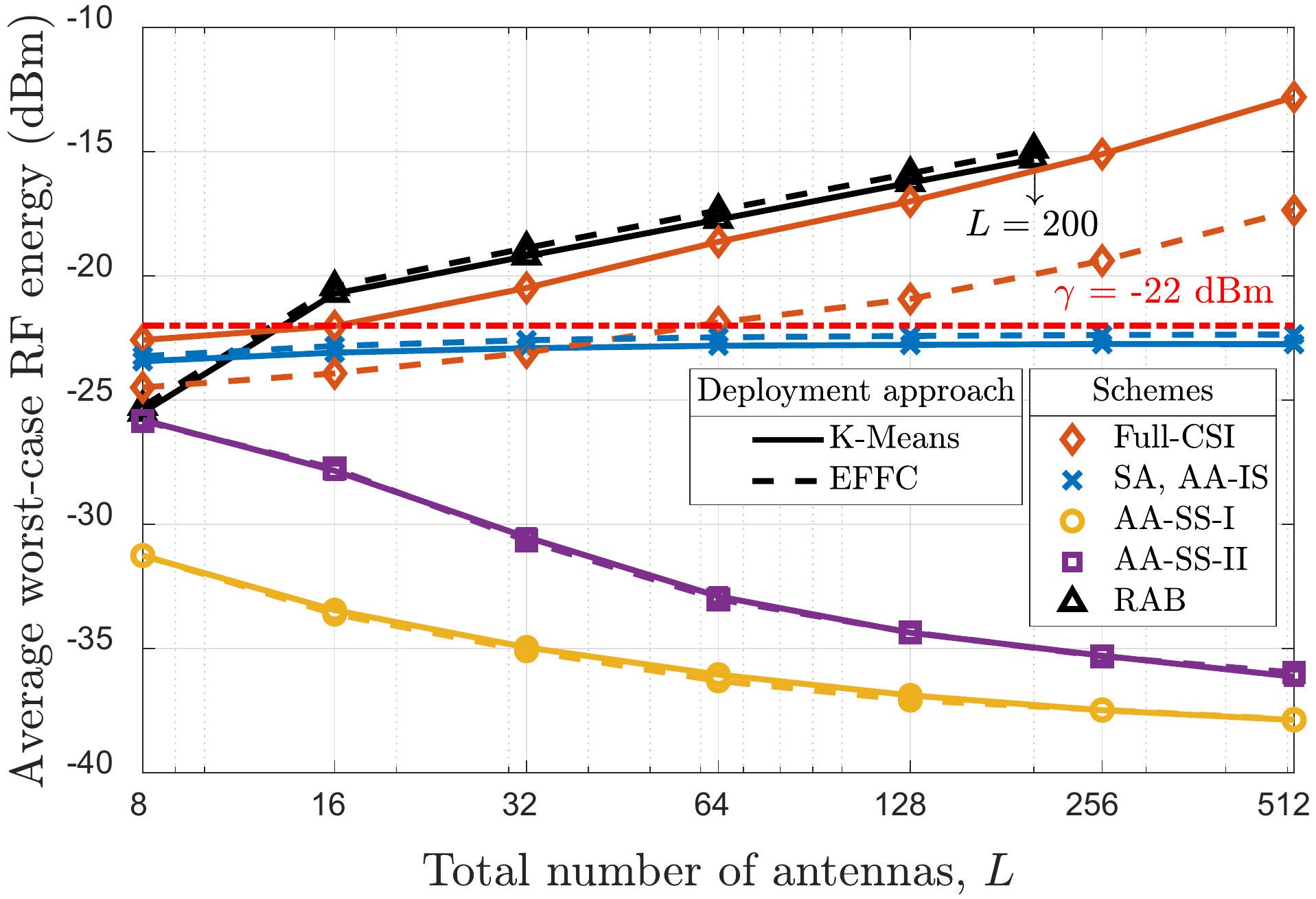}}
    \subfigure[$M=8$]{
    \includegraphics[width=2.3in]{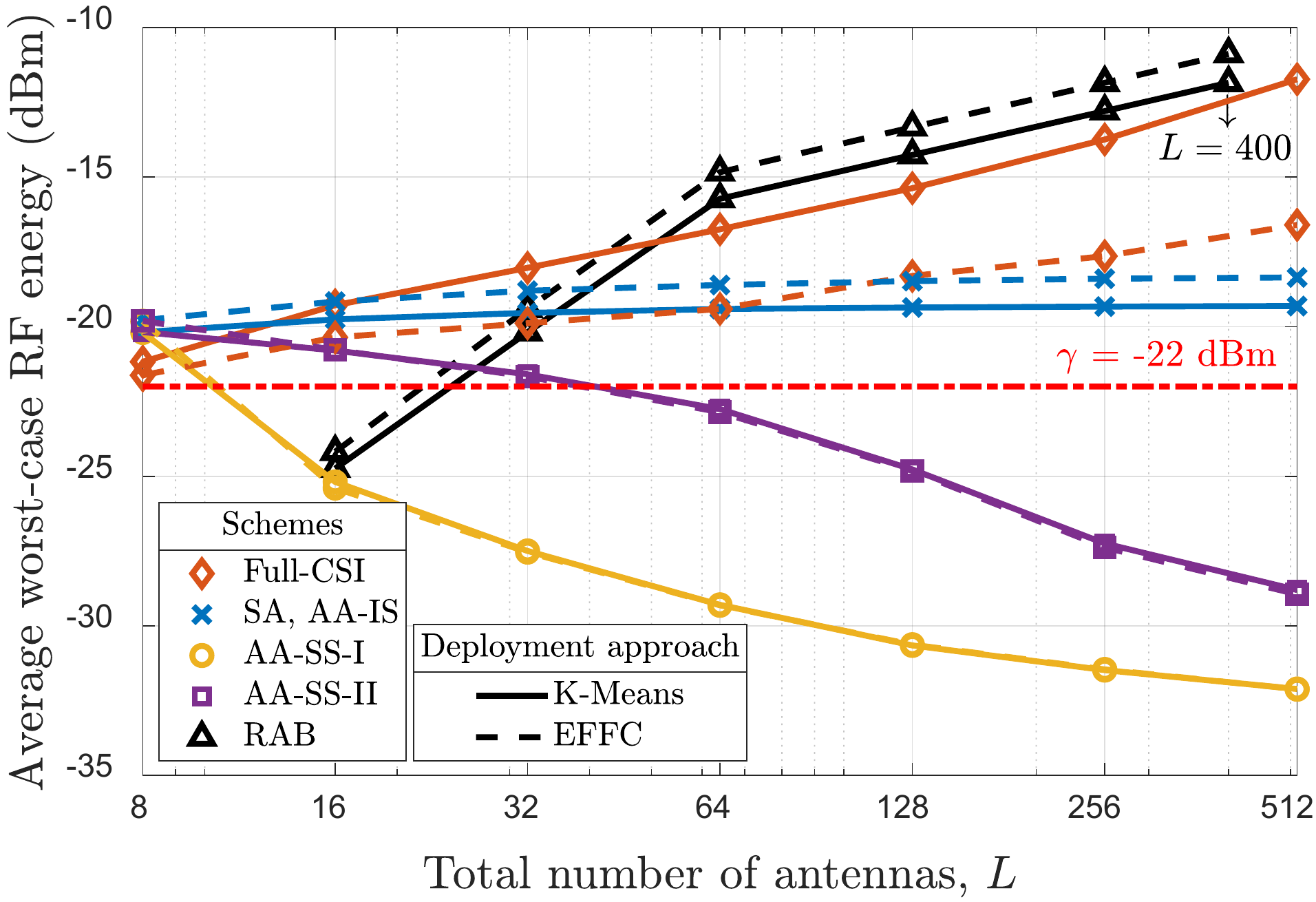}}
	\caption{Average worst-case RF energy available under the ideal system as a function of the total number of antennas $L=MQ$, and  for the full-CSI and CSI-free schemes, number of PBs $M$, and considering the K-Means and EFFC approaches. The red dash-dotted line depicts the EH threshold of $\gamma=-22$~dBm.}
	\label{Ideares}
\end{figure*}
Fig.~\ref{Ideares} depicts the average worst-case RF energy availability delivered to $N=64$ underground EH devices as a function of the total number of antennas $L=MQ$ under the ideal system with $p = P_{budget}$. Herein, the illustrated performance corresponds to the harvester featuring the minimum average RF energy available at the EH circuit’s input among the entire set of underground EH devices. Specifically, Fig.~\ref{Ideares}~(a) displays the performance of a single PB located at the circle center, while Figs.~\ref{Ideares}~(b) and~(c) represent the performance of $M=4$ and $M=8$ PBs, whose positions are set by using the K-Means and EFFC approaches, respectively. Observe that the performance of AA-SS-I and AA-SS-II degrades with an increasing $L$ since a greater number of antennas leads to narrower energy beams. Meanwhile, with a higher $L$, the average RF energy under SA and AA-IS remains almost the same, while that of RAB improves under the ideal system because the smaller the rotation angle improves the probability of powering the whole area more uniformly. Notice that a greater number of rotation steps brings a significant increase in the energy consumption, which will be analyzed in Section~\ref{SecPracticalRes}. Furthermore, the full-CSI strategy exhibits an improvement trend as $L$ increases driven by more flexible and higher-gain beamforming realizations.

Fig.~\ref{Ideares} shows that the average RF energy available at the nodes improves with the number of deployed PBs. This is because the transmit power of the whole system increases while homogenizing the energy distributed in the service area, especially if the PBs are properly deployed. Fig.~\ref{Ideares}~(a) shows that only RAB with $L=50$ antennas and the full-CSI strategy with $L \geq 128$ can guarantee to surpass the EH threshold when a single PB is deployed. Interestingly, RAB outperforms the full-CSI strategy for $L \leq 50$. This is because RAB exploits the mechanical rotation to improve the charging coverage probability, while the full-CSI strategy does not. However, the performance gain is strongly affected by the number of antennas, the number of underground EH devices, and the Rician factor $\kappa$. Note that it is infeasible to obtain the performance of the full-CSI strategy when the number of antennas per PB is $Q=512$, owing to the limitation of the SDP-based solver. As shown in Fig.~\ref{Ideares}~(b), the EH threshold can be overtaken with $M=4$ four-antenna PBs operating with RAB, for which K-Means and EFFC perform alike. When operating the full-CSI strategy, $M=4$ K-Means-based PBs can exceed the EH threshold as $L \geq 32$, which performs significantly better than EFFC.  This is because each PB can specialize the precoder for the underground EH devices assigned to it within the cluster through K-Means. However, for the EFFC approach, the PBs must design the precoder for all devices in the full-CSI strategy owing to the unknown devices' positions. Fig.~\ref{Ideares}~(c) demonstrates that both the CSI-full strategy and the discussed CSI-free schemes can surpass the EH threshold by setting an appropriate $L$ when deploying $M=8$ PBs. Herein, SA and AA-IS have better performance than the other CSI-free schemes as $L \leq 32$, while RAB shows the best performance with $L \geq 64$. Meanwhile, EFFC with SA, AA-IS, and RAB outperforms K-Means in terms of average RF energy. These results reveal that efficient WET performance can be realized without the prior information of the devices' location and that the deployment of distributed CSI-free systems is promising in WPUSNs.    

\subsection{On the Practical CSI-free Distributed System} \label{SecPracticalRes}
\begin{figure*}[t]
	\centering
	\subfigure[$M=1$]{
    \includegraphics[width=2.3in]{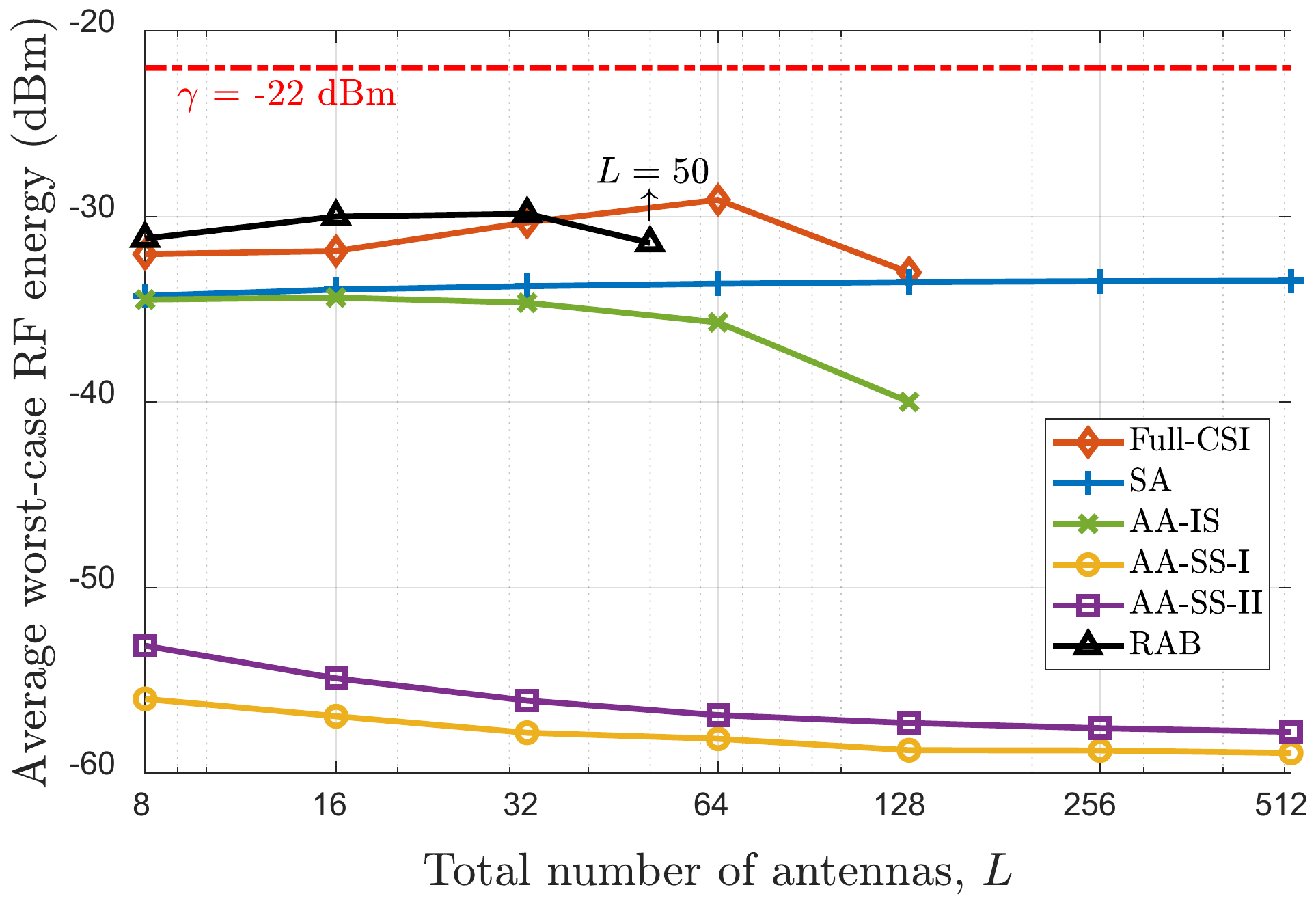}}
    \subfigure[$M=4$]{
    \includegraphics[width=2.3in]{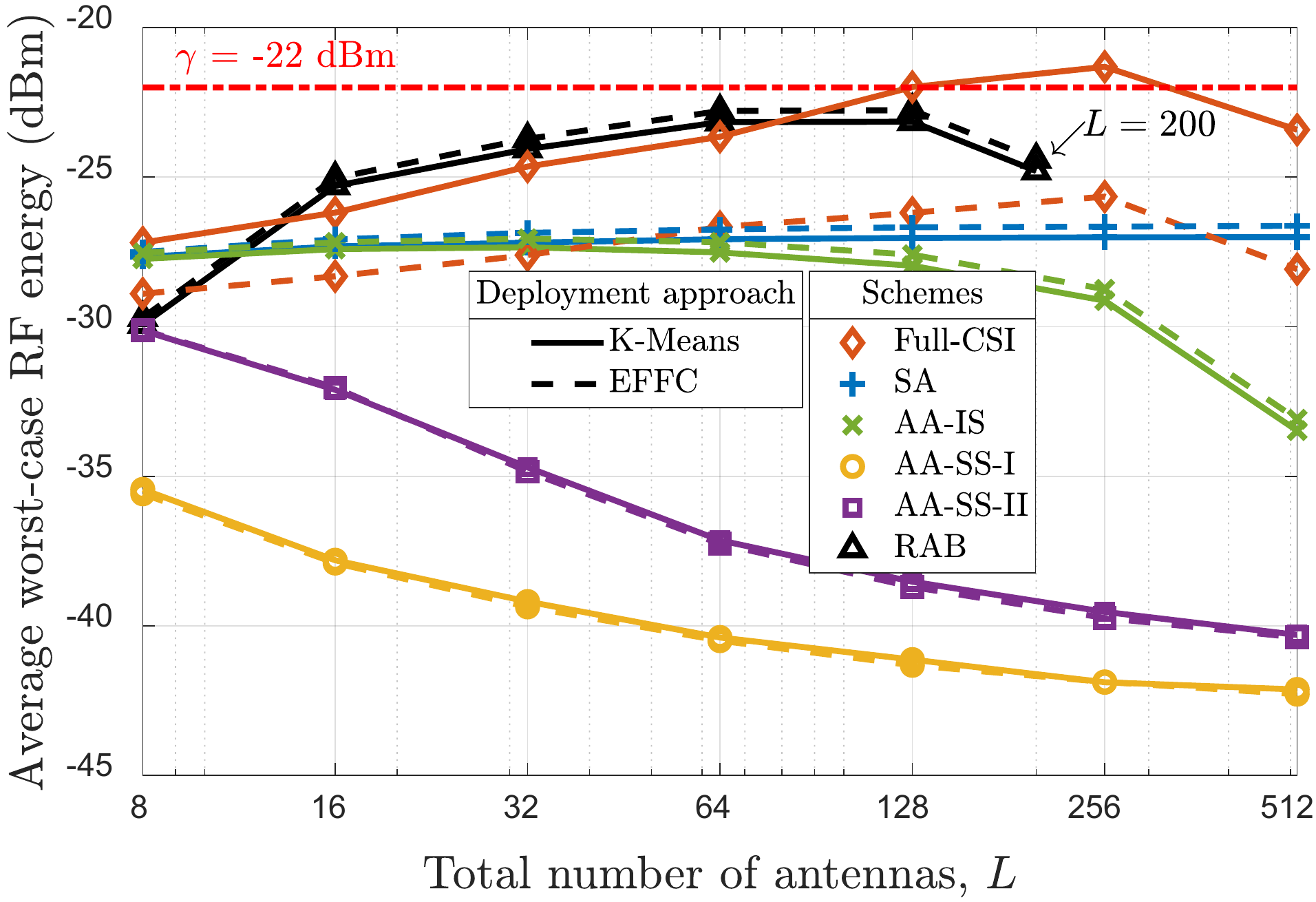}}
    \subfigure[$M=8$]{
    \includegraphics[width=2.3in]{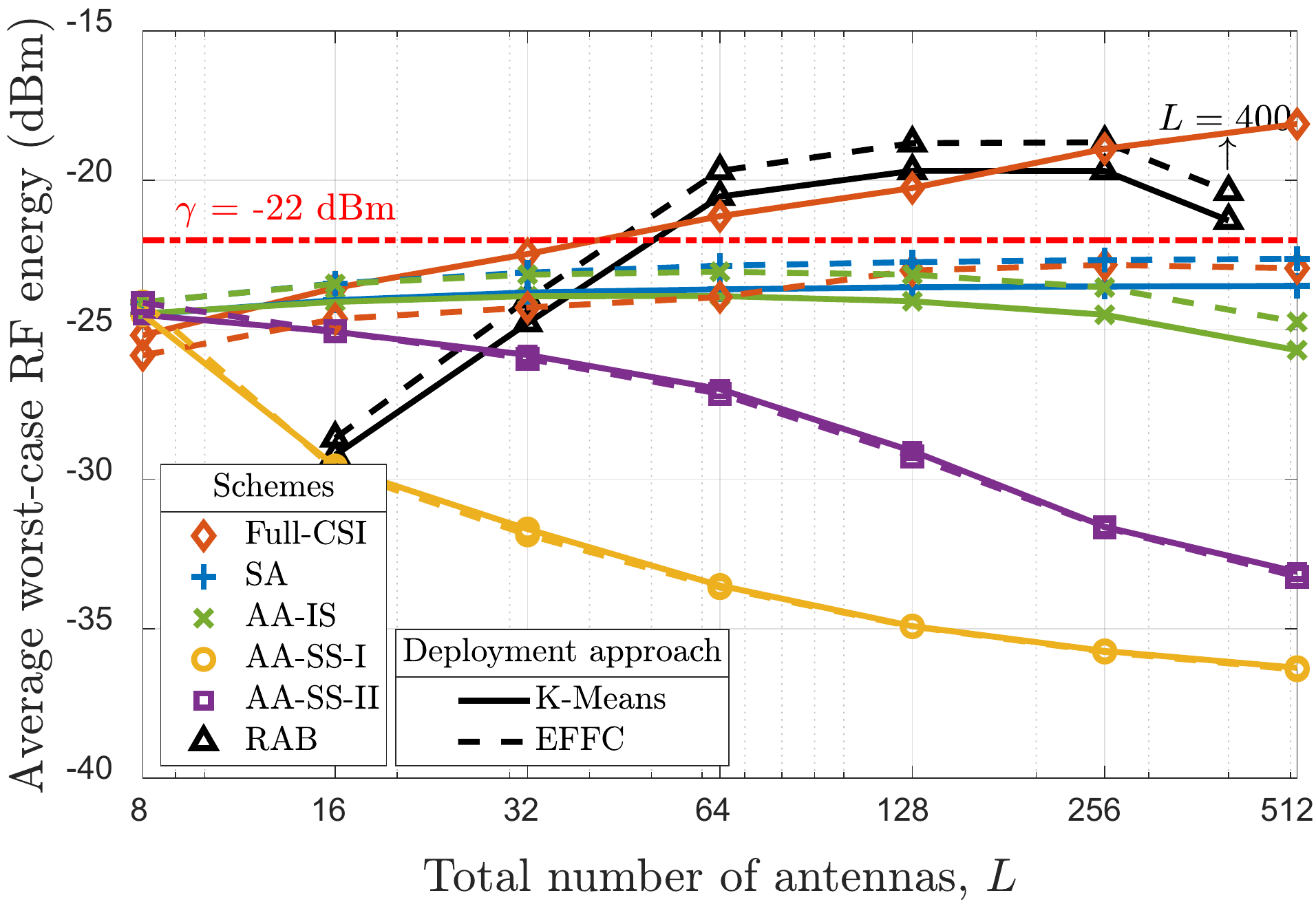}}
	\caption{Average worst-case RF energy available under the practical system as a function of the total number of antennas $L=MQ$, and for the full-CSI and CSI-free schemes, number of PBs $M$, and considering the K-Means and EFFC approaches. The red dash-dotted line depicts the EH threshold of $\gamma=-22$~dBm.}
	\label{Practicalres}
\end{figure*}

Similar to Fig.~\ref{Ideares}, Fig.~\ref{Practicalres} depicts the average worst-case RF energy for a varying number of antennas and PBs, but the difference is that Fig.~\ref{Practicalres} considers the practical system as described in Section~\ref{Secpracticalsys}. As expected, the performance of the practical system degrades compared with the ideal system due to the power consumed by the circuitry and operations. For instance, Fig.~\ref{Practicalres}~(a) shows that the average RF energy under RAB increases with $L$ for $L \leq 32$, and decreases from there on. Such a phenomenon is mostly due to the fact that a larger $L$ boosts the radiation performance of RAB while exacerbating the power consumption of motor's operations as illustrated in~\eqref{RABEnergy1}-\eqref{RABEnerg3}. Similarly, the performance of AA-IS deteriorates with the number of antennas since the increase in RF chains causes higher base-band processing power consumption. Meanwhile, the average RF energy of the full-CSI strategy undergoes an upward trend within $8 \leq L \leq 64$ before subsequently decreasing. Moreover, RAB outperforms the full-CSI strategy when $L \leq 32$. Fig.~\ref{Practicalres}~(b) reveals that the EFFC-based RAB with $L=128$ provides the highest RF energy under the practical CSI-free distributed system with $M=4$ PBs, however it cannot reach the EH threshold. Note that in the EFFC-based RAB scheme, $L=64$ antennas can also realize the similar performance, but are obviously with reduced hardware complexity and operational costs compared to $L=128$ antennas. Despite the K-Means-based full-CSI strategy with $L=256$ can guarantee the EH threshold, it necessitates a substantial amount of time and power for the CSI training and the precoder optimization. Fig.~\ref{Practicalres}~(c) demonstrates that the system's performance can be further improved by placing more PBs, e.g., $M=8$. For instance, the average RF energy of EFFC-based RAB attains $-19$~dBm (which is higher than $\gamma$) when each PB is equipped with $Q=32$ antennas, i.e., $L=256$. In the practical system, SA under EFFC performs better than the other CSI-free schemes under $L \leq 32$ and $L>400$, while the EFFC-based RAB is the winner among all CSI-free schemes as $ 64\leq L \leq 400$. Furthermore, the K-Means-based full-CSI strategy performs better than the discussed CSI-free schemes in the cases of $L=32$ and $512$. According to the practical performance of the aforementioned WET schemes, in Section~\ref{Secunderres}, SA and RAB under the EFFC approach and the K-Means-based full-CSI strategy are adopted in evaluating the system's performance with different VWC and burial depths.     


\subsection{On the Impact of VWC and Burial Depth} \label{Secunderres}
\begin{figure}[t]
    \centering
    \includegraphics[width=3.45in]{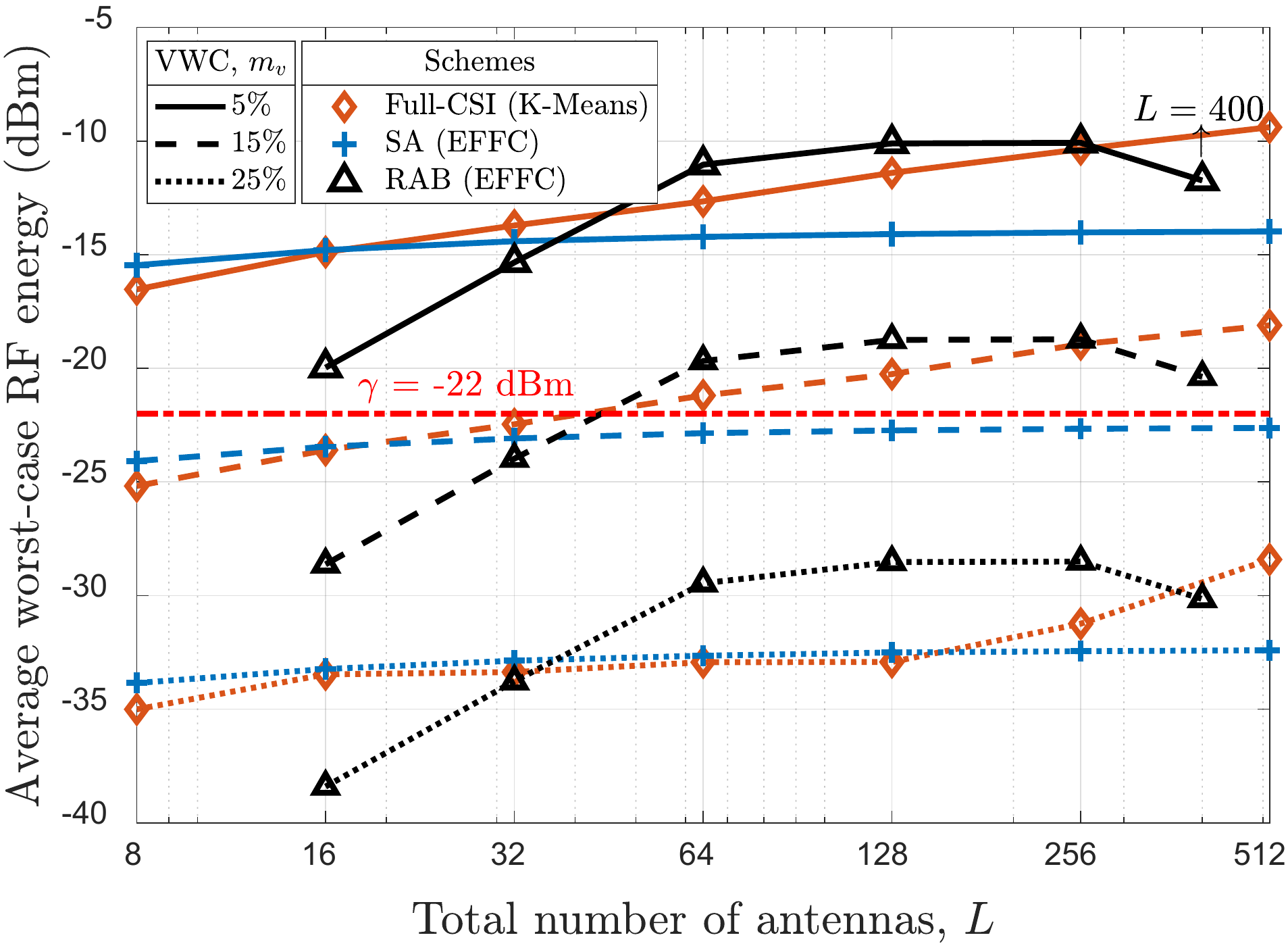}
    \caption{Average worst-case RF energy available under the practical system as a function of the total number of antennas $L=MQ$. We assess the performance of the K-Mean-based full-CSI strategy and the CSI-free schemes (SA and RAB) considering EFFC approach for VWC corresponding to 5\%, 15\%, and 25\% under $M=8$ PBs. The red dash-dotted line depicts the EH threshold of $\gamma=-22$~dBm.}
    \label{VWCRes}
\end{figure}

\begin{figure}[t]
    \centering
    \includegraphics[width=3.45in]{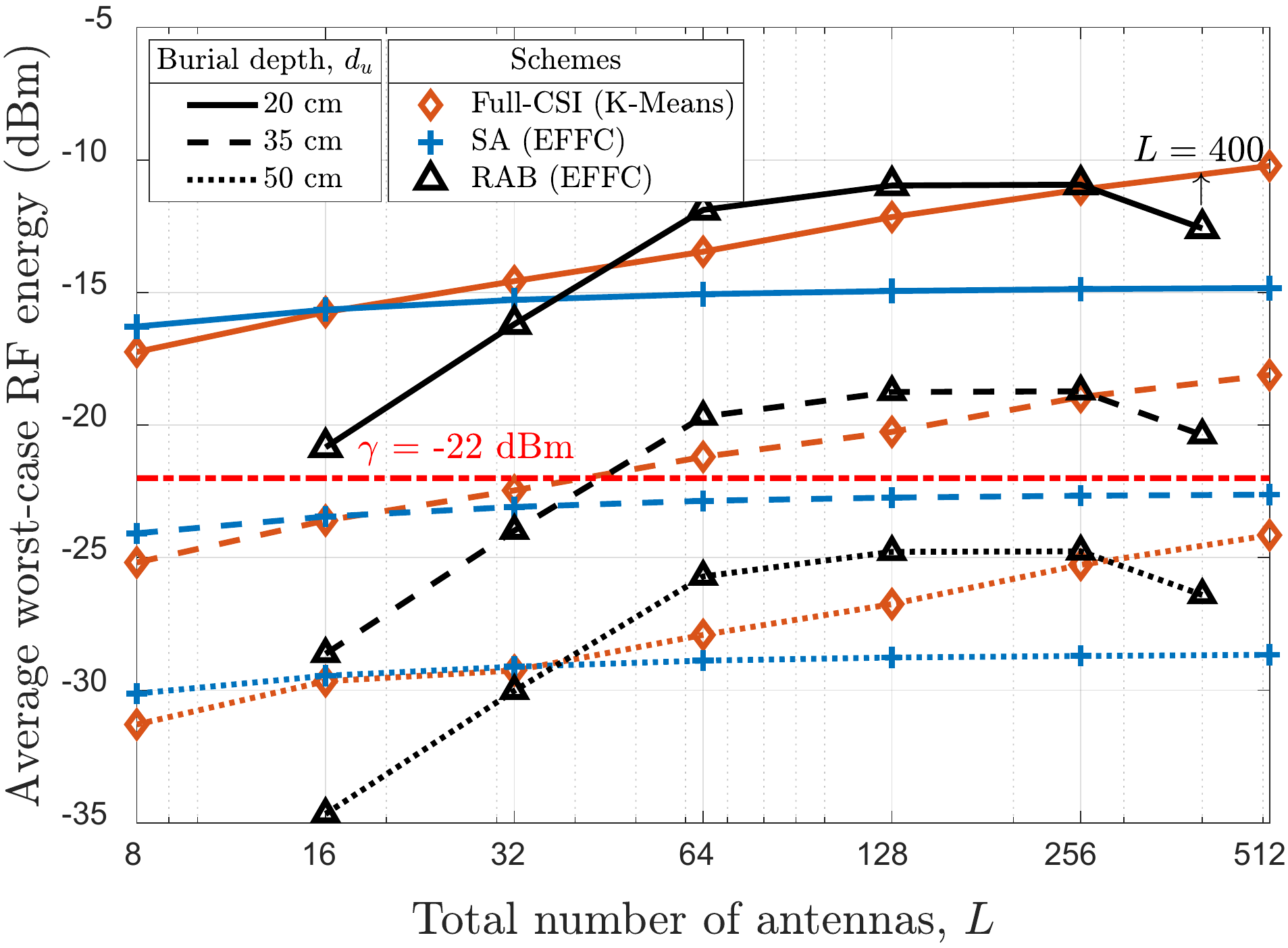}
    \caption{Average worst-case RF energy available under the practical system as a function of the total number of antennas $L=MQ$. We assess the performance of  the K-Mean-based full-CSI strategy and the CSI-free schemes (SA and RAB) considering EFFC approach for burial depths corresponding to 20~cm, 35~cm, and 50~cm under $M=8$ PBs, considering EFFC approach. The red dash-dotted line depicts the EH threshold of $\gamma=-22$~dBm.}
    \label{DepthRes}
\end{figure}

Due to the VWC of soil changing with precipitation or field irrigation, the investigation of the system's performance under different VWC is crucial for the WPUSNs applications with distributed CSI-free schemes. Fig.~\ref{VWCRes} depicts the practical average worst-case RF energy of the K-Mean-based full-CSI strategy and the SA and RAB schemes under EFFC with a varying number of antennas under various VWC values, where $M=8$ PBs are deployed, and the buried depth of underground EH devices is $d_u=35$~cm. As expected, the performance of the distributed WET system deteriorates with VWC, since a larger VWC is a decisive factor for larger soil attenuation which greatly influences the incident RF energy. For instance, the average RF energy of SA and RAB diminishes by around 9~dBm with every 10\% increase in VWC. For $m_v=25\%$, no WET schemes reach the EH threshold. To avoid the pointless power expenditure in the high-VWC underground scenarios, the PBs shall stop the WET operations until the soils get drier. Nevertheless, the results evidence that the full-CSI strategy and RAB with a proper number of antennas can efficiently charge the devices buried at the depth of 35~cm when the VWC is less than 15\%.    

It is of great significance to evaluate the feasibility of the distributed CSI-free system under various burial depths. For instance, in the precision agriculture scenario, the sensors shall be buried at different depths to detect the soil conditions due to the different crop types and root depths. Fig.~\ref{DepthRes} depicts the practical performance of the K-Mean-based full-CSI strategy and the SA and RAB schemes under EFFC as a function of the total number of antennas under different burial depths, where $m_v=15\%$ and $M=8$ PBs are deployed. It is observed that the average RH energy provided by both the full-CSI strategy and the CSI-free schemes degrades by $6\sim8$~dBm when the burial depth increases by 15~cm. As the burial depth is higher than 50~cm, all WET schemes fail to exceed the EH threshold and are not capable of charging devices. Therefore, more PBs and higher transmit power are needed for the deeper WPUSNs scenarios, e.g., underground pipeline monitoring. Furthermore, one can see in~Fig.~\ref{DepthRes} that RAB with $ 64\leq L \leq 400$ antennas can provide the effective WET for massive underground EH devices buried at the depth of 35~cm, while the average RF energy of the full-CSI strategy with $L=512$ can attain $-18$~dBm. Note that 35~cm is a safe and suitable burial depth in smart agriculture applications~\cite{Undergroundfield}.       

\section{Conclusion and Outlook} \label{Conclusion}

WPUSNs are promising for realizing sustainable underground monitoring. Notably, the performance of CSI-based WPUSNs significantly decays with the number of devices due to difficult and costly CSI acquisition procedures, thus motivating the use of CSI-free charging mechanisms. In this work, we extended several recently proposed CSI-free schemes, namely SA, AA-IS, AA-SS, and RAB, to the underground domain, specifically to power massive underground EH devices. Moreover, we proposed a distributed CSI-free system for WPUSNs by exploiting two deployment approaches, i.e., K-Means, and EFFC. Through extensive modeling of a realistic farming scenario under the ideal and practical energy budgets, our numerical results revealed that RAB outperforms the full-CSI strategy under strong LOS conditions and the distributed WET system equipped by RAB with an appropriate number of antennas could efficiently charge massive underground devices. We noticed that, although adopting more antennas enhances the radiation capabilities, it is at the cost of raising the power consumed by the motor operations; therefore, there is an optimum number of transmit antennas for RAB in practical systems. Furthermore, the performance of EFFC, under SA, AA-IS, and RAB, was shown to be slightly better than that of K-Means, demonstrating that efficient WET can be attained even without prior information of devices' positions, which shall help facilitate the deployment of the distributed CSI-free system in WPUSNs. Our results also indicated that the performance of our proposed distributed CSI-free system is strongly affected by two critical underground factors, i.e., VWC and burial depth. All in all, our work confirms the potential feasibility of distributed CSI-free systems in WPUSNs, and offers a guideline for future experimentation and practical deployments to take this novel concept further.

It is worth noting that although we used smart agriculture as the current study case, the proposed distributed CSI-free system can be generalized for other underground applications by adjusting the system parameters. However, there are several open challenges and research directions when merging the CSI-free schemes and WPUSNs. Specifically, 
\begin{enumerate}
\item To realize/enhance the charging directivity toward the underground EH devices while minimizing the loss in the air, it might be interesting to consider directional antenna structures of ULA~\cite{Directive} (and different antenna array typologies~\cite{AntennaWET}) and design CSI-free schemes accordingly. Note that the optimum antenna array implementation shall be specialized for the underground scenarios.

\item Novel mechanisms to intelligently optimize the transmit power of PBs by exploiting the VWC information may be developed. They may enable efficient WET with a lower power supply. Meanwhile, sustainable WPUSNs can be further realized by letting the PBs be powered by ambient energy sources, which shall motivate more investigations on energy scheduling and cooperation protocols. 

\item  Numerous PBs are required to enable the CSI-free schemes for large-scale farms; thus, mounting the PB onto agricultural unmanned aerial vehicles (UAVs) may be an economical way to wirelessly charge the whole area, where the optimal UAV’s trajectory planning shall be studied. 

\item The propagation channel is affected by different underground structures due to soil heterogeneity, sink holes, and subsurface clutters such as stones, rocks, or pebbles. Therefore, further full-wave simulations and field experiments need to be conducted to investigate the WET efficiency of distributed CSI-free systems in such practical environments.


\item This study focuses on the WET phase; however, the uplink wireless information transfer (WIT) is another relevant phase in WPUSNs with many open challenges, especially for a harmonized/optimized WET-WIT coexistence~\cite{CSIfreeTWC}.
\end{enumerate}


\bibliographystyle{IEEEtran}
\bibliography{ref}

\if {0}



 
\vspace{13em}

\begin{IEEEbiography}[{\includegraphics[width=1in,height=1.25in,clip,keepaspectratio]{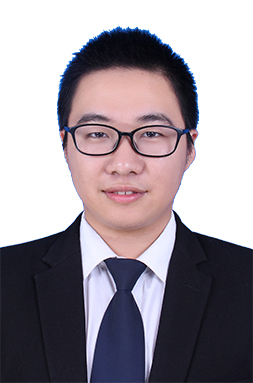}}]{Kaiqiang Lin}
(Student Member, IEEE) received the B.Eng. degree from Fujian Agriculture and Forestry University, Fujian, China, in 2018. He is currently pursuing the Ph.D. degree at the College of Surveying and Geo-Informatics, Tongji University, Shanghai, China. Since November 2021, he is also a visiting Ph.D. student in the Centre for Wireless Communications (CWC), University of Oulu. His current research interests include wireless underground sensor networks, low-power wide-area networks, wireless energy transfer, and sustainable underground monitoring.
\end{IEEEbiography}

\vspace{11pt}
\begin{IEEEbiography}[{\includegraphics[width=1in,height=1.25in,clip,keepaspectratio]{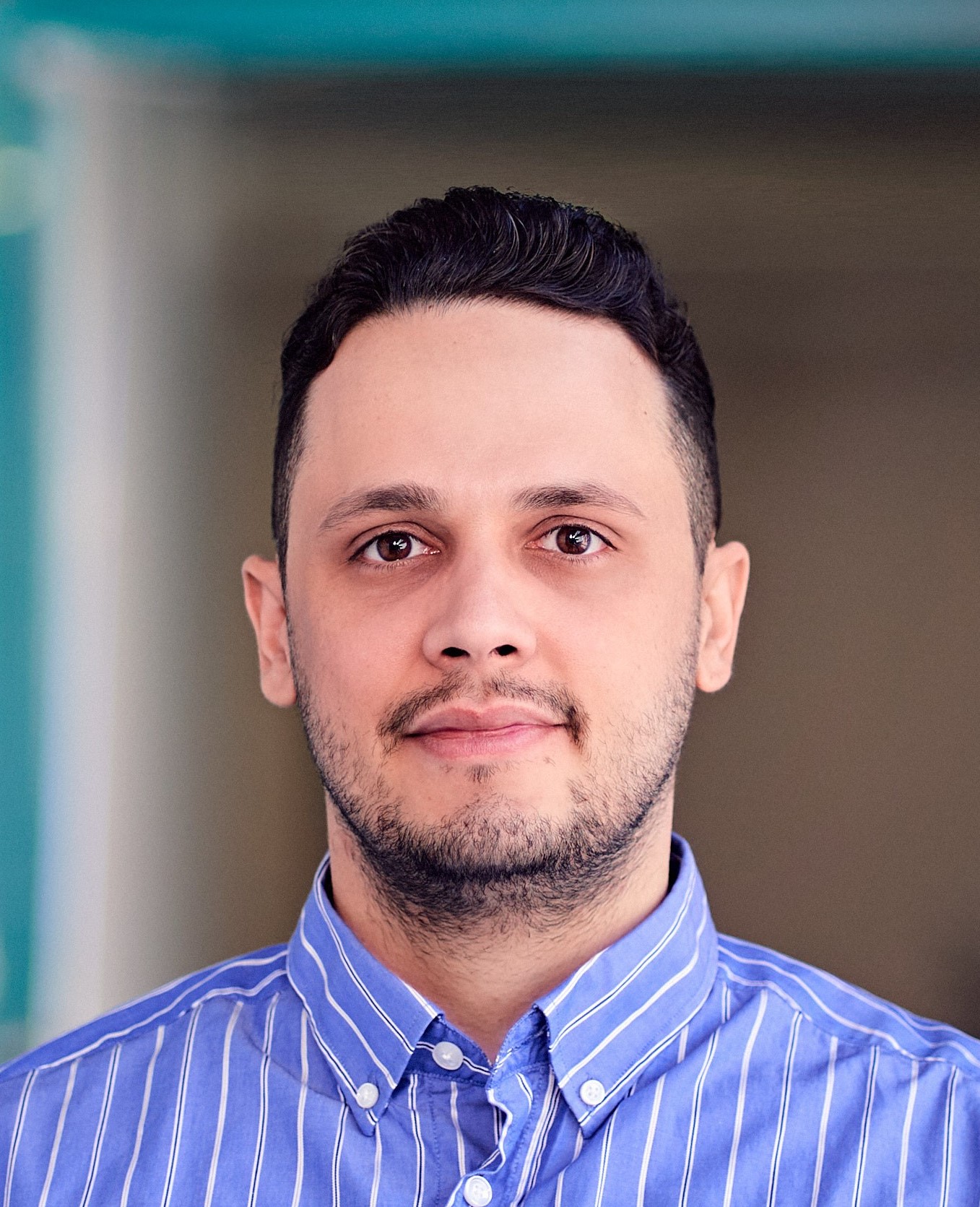}}]{Onel Luis Alcaraz López}
(Member, IEEE) received the B.Sc. (1st class honors, 2013), M.Sc. (2017), and D.Sc. (with distinction, 2020) degree in Electrical Engineering from the Central University of Las Villas (Cuba), the Federal University of Paraná (Brazil), and the University of Oulu (Finland), respectively. From 2013-2015 he served as a specialist in telematics at the Cuban telecommunications company (ETECSA). He is a collaborator to the 2016 Research Award given by the Cuban Academy of Sciences, a co-recipient of the 2019 IEEE European Conference on Networks and Communications (EuCNC) Best Student Paper Award, the recipient of both the 2020 best doctoral thesis award granted by Academic Engineers and Architects in Finland TEK and Tekniska Föreningen i Finland TFiF in 2021 and the 2022 Young Researcher Award in the field of technology in Finland. He is co-author of the book entitled ``Wireless RF Energy Transfer in the Massive IoT era: towards sustainable zero-energy networks'', Wiley, Dec 2021. He currently holds an Assistant Professorship (tenure track) in sustainable wireless communications engineering at the Centre for Wireless Communications (CWC), Oulu, Finland. His research interests include sustainable IoT, energy harvesting, wireless RF energy transfer, wireless connectivity, machine-type communications, and cellular-enabled positioning systems.
\end{IEEEbiography}

\vspace{11pt}
\begin{IEEEbiography}[{\includegraphics[width=1in,height=1.25in,clip,keepaspectratio]{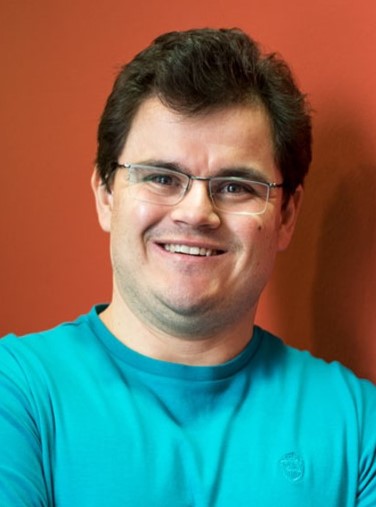}}]{Hirley Alves}
(Member, IEEE) received the B.Sc. and M.Sc. degrees from the Federal University of Technology-Paraná (UTFPR), Brazil, in 2010 and 2011, respectively, in electrical engineering, and the dual D.Sc. Degree from the University of Oulu and UTFPR in 2015. In 2017, he was an Adjunct Professor in machine-type wireless communications with the Centre for Wireless Communications (CWC), University of Oulu, Oulu, Finland. He is an Associate Professor and Head of the Machine-type Wireless Communications Group. He is working on massive connectivity and ultra-reliable low latency communications for future wireless networks, 5GB and 6G, full-duplex communications, and physical-layer security. He leads the URLLC activities for the 6G Flagship Program. He is a co-recipient of the 2017 IEEE International Symposium on Wireless Communications and Systems (ISWCS) Best Student Paper Award and the 2019 IEEE European Conference on Networks and Communications (EuCNC) Best Student Paper Award co-recipient of the 2016 Research Award from the Cuban Academy of Sciences. He has been the organizer, chair, TPC, and tutorial lecturer for several renowned international conferences. He is the General Chair of the ISWCS 2019 and the General Co-Chair of the 1st 6G Summit, Levi 2019, and ISWCS 2020.
\end{IEEEbiography}

\vspace{11pt}
\begin{IEEEbiography}[{\includegraphics[width=1in,height=1.25in,clip,keepaspectratio]{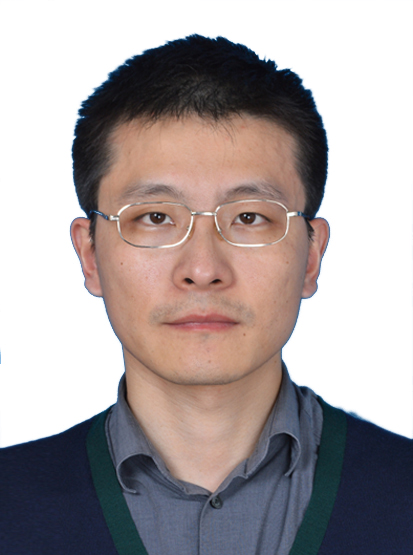}}]{Tong Hao}
(Member, IEEE) received the B.Sc. degree from Nanjing University, Nanjing, China, in 2003, the M.Phil. degree in electronic and electrical engineering from the University of Bath, Bath, U.K., in 2004, and the D.Phil. degree in engineering science from the University of Oxford, Oxford, U.K., in 2009. From 2010 to 2011, he was a Postdoctoral Research Fellow with the University of Birmingham, Birmingham, U.K. From 2012 to 2014, he was a Specialist with the Oil and Gas Sector, General Electric Company, Farnborough, U.K. He is currently a Full Professor with the College of Surveying and Geo-Informatics, Tongji University, Shanghai, China. His current research interests include nondestructive testing using electromagnetic and acoustic technologies, enhanced detection, and sensing of subsurface structure and targets.
\end{IEEEbiography}


\vfill
\fi

\end{document}